\documentclass[a4paper,aps,prd,twocolumn,showkeys,showpacs,nofootinbib]{revtex4}
\usepackage{latexsym}
\usepackage{amsbsy}
\usepackage{mathrsfs}
\usepackage{color}
\usepackage{psfrag}
\usepackage{enumerate}
\usepackage{amsmath,amssymb,calc,amsfonts}
\usepackage{graphicx,calc,epsfig}
\usepackage[multiple]{footmisc}
\usepackage[english]{babel} 
\usepackage{mathtools}
\usepackage{bbm}
\usepackage{accents}
\usepackage{dsfont}
\usepackage{rotating}
\usepackage{bbold}

\def\ut#1{\rlap{\lower1ex\hbox{$\sim$}}#1{}}

\newcommand{\be}{\nopagebreak[3]\begin{equation}}
\newcommand{\ee}{\end{equation}}
\newcommand{\ba}{\nopagebreak[3]\begin{eqnarray}}
\newcommand{\ea}{\end{eqnarray}}
\DeclareFontFamily{U}{rsfs}{}         
\DeclareFontShape{U}{rsfs}{m}{n}{<5> rsfs5 <6><7> rsfs7          
  <8><9><10><10.95><12><14.4><17.28><20.74><24.88> rsfs10}{}     
\DeclareMathAlphabet{\mathfs}{U}{rsfs}{m}{n}                     
                               
\newcommand{\tr}{\mathrm{tr}}

\def\pb#1{\rlap{\lower1.5ex\hbox{$\longleftarrow$}}{#1}}
\def\dpb#1{\rlap{\lower1.5ex\hbox{$\Longleftarrow$}}{#1}}
\def\spb#1{\rlap{\lower1.5ex\hbox{$\leftarrow$}}{#1}}
\def\sdpb#1{\rlap{\lower1.5ex\hbox{$\Leftarrow$}}{#1}}

\definecolor{blue}{rgb}{0,0,1}
\definecolor{green}{rgb}{0,1,0}
\definecolor{red}{rgb}{1,0,0}
\definecolor{vio}{rgb}{1,0,1}
\definecolor{ama}{rgb}{1,1,0}
\usepackage[all]{xy}
\xyoption{knot}
\xyoption{frame}
\xyoption{arc}
\xyoption{curve}
\xyoption{poly}
\usepackage{hyperref}
\hypersetup{
  colorlinks   = true, 
  urlcolor     = blue, 
  linkcolor    = blue, 
  citecolor   =  red 
}
\def\be{\begin{equation}}
\def\ee{\end{equation}}
\def\ba{\begin{eqnarray}}
\def\ea{\end{eqnarray}}

\begin{document}

\begin{flushleft}
KCL-PH-TH/2016-43
\end{flushleft}

\title{Impact of nonlinear effective interactions on GFT quantum gravity condensates}

\date{\today}

\author{Andreas G. A. Pithis}\email{andreas.pithis@kcl.ac.uk}
\author{Mairi Sakellariadou}\email{mairi.sakellariadou@kcl.ac.uk}
\affiliation{Department of Physics, King's College London, University of London, Strand, London, WC2R 2LS, U.K., EU}
\author{Petar Tomov}\email{tomov@mathematik.hu-berlin.de}
\affiliation{Institut f\"ur Physik, Humboldt-Universit\"at zu Berlin, 12489 Berlin, Germany, EU}

\thanks{}

\begin{abstract}
We present the numerical analysis of effectively interacting Group Field Theory (GFT) models in the context of the GFT quantum gravity condensate analog of the Gross-Pitaevskii equation for real Bose-Einstein condensates including combinatorially local interaction terms. Thus we go beyond the usually considered construction for free models.

More precisely, considering such interactions in a weak regime, we find solutions for which the expectation value of the number operator $N$ is finite, as in the free case. When tuning the interaction to the strongly nonlinear regime, however, we obtain solutions for which $N$ grows and eventually blows up, which is reminiscent of what one observes for real Bose-Einstein condensates, where a strong interaction regime can only be realized at high density. This behavior suggests the breakdown of the Bogoliubov ansatz for quantum gravity condensates and the need for non-Fock representations to describe the system when the condensate constituents are strongly correlated.
 
Furthermore, we study the expectation values of certain geometric operators imported from Loop Quantum Gravity in the free and interacting cases. In particular, computing solutions around the nontrivial minima of the interaction potentials, one finds, already in the weakly interacting case, a nonvanishing condensate population for which the spectra are dominated by the lowest nontrivial configuration of the quantum geometry. This result indicates that the condensate may indeed consist of many smallest building blocks giving rise to an effectively continuous geometry, thus suggesting the interpretation of the condensate phase to correspond to a geometric phase.

\end{abstract}


\maketitle

\section{Introduction}
The most difficult problem for all quantum gravity approaches using discrete and quantum pregeometric structures is the recovery of continuum spacetime, its geometry, diffeomorphism invariance and General Relativity as an effective description for the dynamics of the geometry in an appropriate limit. It has been suggested, a possible way of how continuum spacetime and geometry could emerge from a quantum gravity substratum in such theories is by means of at least one phase transition from a discrete pregeometric to a continuum geometric phase. One refers to such a process as "geometrogenesis" \cite{Geometrogenesis}. A particular representative in this class of approaches where such a scenario has been proposed is Group Field Theory (GFT) \cite{GFT} where one tries to identify the continuum geometric phase to a condensate phase of the underlying quantum gravity system \cite{GFTGeometrogenesis} with a tentative cosmological interpretation \cite{GFC,GFCExample,GFCOthers,GFCEmergentFriedmann,GFCEmergentFriedmann2,GFClowspin, GFCEmergentFriedmann3,GFCReview}. 

GFTs are Quantum Field Theories (QFT) defined over group manifolds and are characterized by their combinatorially nonlocal interaction terms. In the perturbative expansion it becomes apparent that the Feynman diagrams of the theory are dual to cellular complexes because of this particular nonlocality. Depending on the details of the Feynman amplitudes, the sum over the cellular complexes can be interpreted as a possible discrete definition of the covariant path integral for $4d$ quantum gravity. The reason for this is that beyond the combinatorial details, GFT Feynman graphs can be dressed by group theoretic data of which the function is to encode geometric information corresponding exactly to the elementary variables of Loop Quantum Gravity (LQG) \cite{LQG}. Using this, it can be shown that GFTs provide a formal and complete definition of spin foam models which give a path integral formulation for LQG \cite{SF,GFTSF}. In case the GFTs possess a discrete geometric interpretation, it is also possible to manifestly relate their partition functions to (noncommutative) simplicial quantum gravity path integrals \cite{GFTSG}.

In order to understand the nonperturbative properties of particular GFT models, the application of Functional Renormalization Group (FRG) techniques is needed \cite{FRG}. In general, these techniques provide the most powerful theoretical description of thermodynamic phases by means of a coarse graining operation that progressively eliminates short scale fluctuations. Their successful application to matrix models of $2d$ quantum gravity \cite{MMs,MMFRG} serves as an example for the adaption to GFT models which has recently been very actively pursued \cite{TGFTFRG,GFTRGReview}. In this way, the FRG methods enable one to study the consistency of GFT models, analyze their continuum limit, chart their phase structure and investigate the possible occurrence of phase transitions.

More precisely, standard FRG methodology has been applied to a couple of models from a class of group field theories called tensorial GFTs, for which one requires the fields to possess tensorial properties under a change of basis. The common features of the models analyzed so far are a nontrivial kinetic term of Laplacian type and a quartic combinatorially nonlocal interaction. However, they differ firstly in the size of the rank, secondly in whether gauge invariance is imposed or not, and thirdly in the compactness or noncompactness of the used group manifold. Remarkably, all these models are shown to be asymptotically free in the UV limit which is deeply rooted in the combinatorial nonlocality of the interaction \cite{TMRG}. Furthermore, signs for a phase transition separating a symmetric from a broken/condensate phase were found as the mass parameter $\mu$ tends to negative values in the IR limit analogous to a Wilson-Fisher fixed point in the corresponding local QFT. To corroborate the existence of such a phase transition, among others, the theory has to be studied around the newly assumed ground state by means of a mean field analysis as noticed in Refs. \cite{TGFTFRG,GFTRGReview}. One way to check this would amount to finding solutions to the classical equation of motion in a saddle point approximation of the path integral. 

The possible occurrence of a phase transition in such systems is highly interesting, since it has been suggested that phase transitions from a symmetric to a condensate phase in GFT models for $4d$ quantum gravity could be a realization of the above-mentioned geometrogenesis scenario. In such a setting, a pregeometric discrete phase, given by an appropriate microscopic GFT model, passes through a phase transition into a  continuum geometric phase, the dynamics of which is in turn described by a corresponding effective action. The phase transition would then correspond to a RG flow fixed point and could be interpreted as the condensation of discrete spacetime building blocks \cite{Geometrogenesis,GFTGeometrogenesis}. 

So far, however, the mentioned FRG results for tensorial GFTs can only lend indirect support to the geometrogenesis hypothesis, since a full geometric interpretation of such models is currently lacking. To realize such a hypothesis in this context, one would have to proceed toward a GFT model enriched with additional geometric data and an available simplicial quantum gravity interpretation that is closely linked to LQG. The application of FRG methods to such a model with a combinatorially nonlocal simplicial interaction term would be needed to give an accurate account of the phase structure of the system. The hope is that studying its renormalization group flow will reveal an IR fixed point which marks the phase transition into a condensate phase ideally corresponding to a continuum geometric phase. Hence, the aim is to gradually increase the sophistication of the studied toy models to rigorously underpin the GFT condensate assumption and connect it to the geometrogenesis hypothesis \cite{GFTGeometrogenesis}.

In this context, the basic aim of GFT Condensate Cosmology (GFTCC) is to derive an effective dynamics for the GFT condensate states directly from the microscopic GFT quantum dynamics using mean field theoretic considerations, and consequently to extract a cosmological interpretation from them \cite{GFC,GFCExample,GFCOthers,GFCEmergentFriedmann,GFCEmergentFriedmann2,GFClowspin,GFCEmergentFriedmann3,GFCReview}. The central assumption of GFTCC is that the possible continuum geometric phase of a particular GFT model is ideally approximated by a condensate state which is suitable to describe spatially homogeneous universes. The mean field theoretic considerations used so far in Refs. \cite{GFC,GFCExample,GFCOthers,GFCEmergentFriedmann,GFCEmergentFriedmann2,GFClowspin,GFCEmergentFriedmann3,GFCReview} to give an effective description of the condensate phase and its dynamics use techniques which are strongly reminiscent of those employed to study the Gross-Pitaevskii equation for, at most, weakly interacting Bose-Einstein condensates \cite{BECs,BogoliubovAnsatz}. 

In this article, we will go beyond the analysis of free GFTCC models and investigate the effect of combinatorially local interaction terms (pseudopotentials) for a gauge invariant model with Laplacian kinetic term, the mean field analysis of which was started for the free case in an isotropic restriction\footnote{The isotropic restriction employed in this article differs from the one used in Ref. \cite{GFCEmergentFriedmann} which renders the interaction term local in the spin label.} \cite{GFCExample}. We will further elaborate the results of the free model in this restriction studying the expectation values of certain geometric operators. We note that no additional massless scalar field is added to study the evolution of the system in relational terms as in Refs.  \cite{GFCEmergentFriedmann,GFCEmergentFriedmann2,GFClowspin,GFCEmergentFriedmann3}. We choose local interactions for pure practical reasons: in this way the equations of motion take a particularly simple nonlinear form and to solve them, we employ numerical methods. Despite the fact that from a physics viewpoint these models appear as somewhat artificial because such interactions lack a proper discrete geometric interpretation, they have nevertheless a practical utility as simplified versions of more complicated ones, and bring us nearer to the physics which we want to probe. One might also speculate that the local effective interactions between the condensate constituents could only be valid on length scales where the true microscopic details of the interaction, namely the combinatorial nonlocality, are irrelevant. Ultimately, rigorous RG arguments will have the last word on how such terms can or cannot be derived from the fundamental theory; however by adopting a phenomenological point of view, the analysis of the effect of such pseudopotentials might prove useful to clarify the map between the microscopic and effective macroscopic regimes of the theory.

For a particular choice of the signs of the free parameters in the GFT action, we find in the isotropic restriction solutions which (i) are consistent with the condensate ansatz, (ii) are normalizable with respect to the Fock space measure in the weakly nonlinear regime, and (iii) obey a specific condition of which the fulfillment is required for the interpretation in terms of continuous smooth manifolds. For such solutions we study the effect of the interactions onto the expectation values of certain operators needed for their further geometric interpretation. We repeat this analysis for solutions to the equations of motion around the nontrivial minima of the used effective potentials and find that the expectation values of the geometric operators are clearly dominated by low spin modes. In this sense, such solutions can be interpreted as giving rise to an effectively continuous geometry. Moreover, we discuss the consequences of the interactions in the strongly nonlinear regime, where solutions generally lose their normalizability with respect to the Fock space measure and thus can be interpreted as corresponding to non-Fock representations of the canonical commutation relations. 

To this aim, the article is organized as follows. In the first part of the second section \ref{GFT} we review the GFT approach to quantum gravity from the classical and quantum perspective. We then motivate its quantum cosmology spinoff called GFTCC in subsection \ref{GFC}. The presentation is kept rather short to motivate the essential concepts needed to follow the analysis presented later on. We invite the reader familiar with these concepts to proceed directly to the third section \ref{GFCInteractions} where we analyze in detail the properties of the free and interacting solutions in subsections \ref{subsectionfree}, \ref{subsectioninteractionssetup},\ref{subsectioninteractionsperturbation} and \ref{subsectioninteractionslocalminima}, respectively. In the fourth section \ref{Discussion} we summarize our results, discuss limitations of our analysis and propose further studies. 

In the appendices \ref{AppendixA}, \ref{AppendixB} and \ref{AppendixC} we supplement the main sections of this article by discussing the notions of noncommutative Fourier transform, non-Fock representations and non-Fock coherent states, respectively, needed to allow for a better understanding of the obtained results.

\section{Group Field Theory and Group Field Theory Condensate Cosmology}\label{GFTGFC}

\subsection{Group Field Theory}\label{GFT}
GFTs represent a particular class of QFTs which aim at generalizing matrix models for $2d$ quantum gravity to higher dimensions. The fields of GFT live on group manifolds $G$ or dually on their associated Lie algebras $\mathfrak{g}$. For quantum gravity intended models, $G$ is interpreted as the local gauge group of gravity.\footnote{Typically, one chooses $G=\textrm{Spin}(4)$, $\textrm{SL}(2,\mathbb{C})$ or $G=\textrm{SU}(2)$. The last is the gauge group of Ashtekar-Barbero gravity lying at the heart of canonical LQG.} The essential idea is that all data encoded in the fields are solely of combinatorial and algebraic nature thus rendering GFT into a manifestly background independent and generally covariant field theoretic framework \cite{GFT}.

In the following, we introduce aspects of this approach in a shortened manner which are needed for its application to GFT condensate cosmology in the remainder of this article.

\subsubsection{Classical theory}
The classical field theory is specified by choosing a type of field and an action dictating its dynamics. Most generally, we consider the complex-valued scalar field $\varphi$ living on $d$ copies of the Lie group $G$, i.e.,
\be
\varphi(g_{I}):G^d\to\mathbb{C}
\ee
with $I=1,...,d$. The group elements $g_I$ are parallel transports $\mathcal{P}e^{i\int_{e_{I}}A}$ associated to $d$ links $e_I$ and $A$ denotes a gravitational connection $1$-form.

Importantly, one demands the invariance under the right diagonal action of $G$ on $G^d$, i.e.,
\be\label{gaugeinvariance}
\varphi(g_1h...,,g_dh)=\varphi(g_1,...,g_d),~~~\forall h\in G
\ee
which is a way to guarantee that the parallel transports, emanating from a vertex and terminating at the end point of their respective links $e_I$, only encode gauge invariant data.

For compact $G$ the action is given by
\be\label{classicalaction}
S[\varphi,\bar{\varphi}]=\int_G (dg)^d \int_G (dg')^d \bar{\varphi}(g_I)\mathcal{K}(g_I,g_{I}')\varphi(g_I')+\mathcal{V},
\ee
where $dg$ stands for the normalized Haar measure on $G$. The symbol $\mathcal{K}$ denotes the kinetic kernel and $\mathcal{V}=\mathcal{V}[\varphi,\bar{\varphi}]$ is a nonlinear and in general nonlocal interaction potential. Choices of $\mathcal{K}$, $\mathcal{V}$, $d$ and $G$ define a specific model. The classical equation of motion is then given by
\be
\int (dg')^d~\mathcal{K}(g_I,g'_I)\varphi(g'_I)+\frac{\delta \mathcal{V}}{\delta \bar{\varphi}(g_I)}=0.
\ee

\subsubsection{Quantum theory: path integral}
The quantum theory is defined by the partition function $Z_{GFT}$. If we write a more general interaction term as a sum of polynomials of degree $n$, i.e. $\mathcal{V}=\sum_n\lambda_n \mathcal{V}_n$, the path integral becomes
\be\label{pathintegral}
Z_{GFT}=\int[\mathcal{D}\varphi][\mathcal{D}\bar{\varphi}]e^{-S[\varphi,\bar{\varphi}]}=\sum_{\Gamma}\frac{\prod_n\lambda_n^{N_n(\Gamma)}}{\textrm{Aut}(\Gamma)}\mathcal{A}_{\Gamma}
\ee
in the perturbative expansion in terms of the coupling constants $\lambda_n$. The Feynman diagrams are denoted by $\Gamma$, $\textrm{Aut}(\Gamma)$ is the order of their automorphism group, $N_n(\Gamma)$ denotes the number of interaction vertices of type $n$, and $\mathcal{A}_{\Gamma}$ is the Feynman amplitude. Crucially, field arguments in $\mathcal{V}$ are related to each other in a specific combinatorially nonlocal pattern which correlates fields among each other just through some of their arguments. This model-specific combinatorial nonlocality implies that the GFT Feynman diagrams are dual to cellular complexes of arbitrary topology \cite{GFT}.

In view of constructing a partition function for $4d$ quantum gravity, one starts with the GFT quantization of the Ooguri model \cite{Ooguri}, a topological BF-theory, which is based on a real field with $G=\textrm{Spin}(4)$ or $\textrm{SL}(2,\mathbb{C})$ and its Feynman diagrams are dual to simplicial complexes. If $d$ is chosen to equal the dimension of the spacetime under construction, the fields are interpreted as $(d-1)$-simplices. The $d$ arguments of the fields are then associated to their $(d-2)$-faces. In this case, a particular type of interaction $\mathcal{V}$ describes how $d+1$ of these simplices are glued together across their faces to constitute the boundary of a $d$-simplex. Finally, the kinetic operator $\mathcal{K}$ dictates how to glue together two such $d$-simplices across a shared $(d-1)$-simplex. 

In particular, for the case we are aiming at, the (right-invariant) field is defined over $d=4$ copies of $G$ and corresponds to a quantum tetrahedron or equally a $3$-simplex, a choice we further motivate in appendix \ref{AppendixA}. For the construction of the corresponding simplicial path integral, the interaction term has five copies of the field. Their arguments are paired in a particular way to form a $4$-simplex, given by
\be
\mathcal{V}=\frac{\lambda}{5!}\int (dg)^{10}~\varphi_{1234}\varphi_{4567}\varphi_{7389}\varphi_{96210}\varphi_{10851}
\ee
with $\varphi(g_1,g_2,g_3,g_4)\equiv \varphi_{1234}$ etc. The kinetic term of the action with kernel $\mathcal{K}(g_I,g'_I)=\delta(g'_Ig_{I}^{-1})$ is specified by
\be
K=\frac{1}{2}\int (dg)^{4}~\varphi_{1234}^2.
\ee
The data given so far does not yet permit the reconstruction of a unique geometry for the simplicial complex. In a second step, one has to impose restrictions which reduce the nongeometric quantum theory to the gravitational sector.

This can be substantiated by invoking the correspondence between GFT and spin foam models. Indeed, any GFT model defines in its perturbative expansion a spin foam model \cite{GFT,GFTSF}. One can then show, that GFTs based on the Ooguri model, may provide a covariant QFT formulation of the dynamics of LQG. In the latter, boundary spin network states correspond to discrete quantum $3$-geometries \cite{LQG} and transition amplitudes in between two such boundary states are given by appropriate spin foam amplitudes \cite{SF}. A concrete strategy to construct gravitational spin foam models is to start with a spin foam quantization of the topological BF-theory which is equivalent to setting up its discrete path integral. Importantly, it is then turned into a gravitational theory by imposing so-called \textit{simplicity constraints}. These restrict the data dressing the spin foam model such that it becomes equivalent to a discrete path integral for Plebanski gravity. Moreover, the constraints allow one to establish the link to LQG by restricting the group $G$ to $\mathrm{SU}(2)$ \cite{SFSC}. 

It is precisely in this way, that each so-constructed spin foam amplitude corresponds to a discrete spacetime history interpolating in between the boundary configurations and is thus identical to a restricted GFT Feynman amplitude. Therefore, the sum over Feynman diagrams given by Eq. (\ref{pathintegral}) can be rewritten as a sum over diagrams dual to simplicial complexes decorated with quantum geometric data which clarifies how the GFT partition function can be intuitively understood to encode the sum-over-histories for $4d$ quantum gravity.\footnote{This reasoning could be generalized in different ways. Firstly, the discussion of the case with noncompact group $G$ can be found in Ref. \cite{GFC}. Secondly, when remaining faithful to simplicial building blocks, one could e.g. consider higher interaction terms which also allow for an interpretation in terms of regular simplicial $4$-polytopes. Finally, it is in principle possible to go beyond the choice of simplicial building blocks and define GFTs which are fully compatible with the combinatorics of LQG. Within this theory, quantum states of the $3$-geometry are defined on boundary graphs with vertices of arbitrary valence. These correspond to general polyhedra and not merely to $3$-simplices \cite{GFT4All}.}

\subsubsection{Quantum theory: 2nd quantized framework}
Motivated by the roots of GFT in LQG, it is possible to construct a $2$nd quantized Fock space reformulation of the kinematical Hilbert space of LQG of which the states describe discrete quantum $3$-geometries. 
The construction is closely analogous to the one known from ordinary nonrelativistic QFTs \cite{GFTLQG,GFTReview}. In a nutshell, the construction leads to the reinterpretation of spin network vertices as fundamental quanta which are created or annihilated by the field operators of GFT. Pictorially seen, exciting a GFT quantum creates an atom of space or a \textit{choron} and thus GFTs are not QFTs on space but of space itself.

To start with, the GFT Fock space constitutes itself from a fundamental single-particle Hilbert space $\mathcal{H}_v=L^2(G^d)$ 
\be
\mathcal{F}(\mathcal{H}_v)=\bigoplus_{N=0}^{\infty}\textrm{sym}\bigl(\otimes_{i=1}^N\mathcal{H}_v^{(i)}\bigr).
\ee
The symmetrization with respect to the permutation group $S_N$ is chosen to account for the choice of bosonic statistics of the field operators and pivotal for the idea of reinterpreting spacetime as a Bose-Einstein condensate (BEC). 
$\mathcal{H}_v$ is the space of states of a GFT quantum. For $G=\textrm{SU}(2)$ and the imposition of gauge invariance as in Eq. (\ref{gaugeinvariance}), a state represents an open LQG spin network vertex or its dual quantum polyhedron.\footnote{This also holds true for $G=\textrm{SL}(2,\mathbb{C})$ and $G=\textrm{Spin}(4)$ when gauge invariance and simplicity constraints are properly imposed.} In the simplicial context, when $d=4$, a GFT quantum corresponds to a quantum tetrahedron, the Hilbert space of which is 
\be\label{HilbertspaceTetrahedron}
\mathcal{H}_v=L^2(G^4/G)\cong\bigoplus_{J_{i}\in\frac{\mathbb{N}}{2}}\textrm{Inv}\bigl(\otimes_{i=1}^4\mathcal{H}^{J_{i}}\bigr),
\ee
with $\mathcal{H}^{J_{i}}$ denoting the Hilbert space of an irreducible unitary representation of $G=\textrm{SU}(2)$.

In this picture, the no-space state in $\mathcal{F}(\mathcal{H}_v)$ is devoid of any topological and quantum geometric information. It corresponds to the Fock vacuum $|\emptyset\rangle$ defined by
\be
\hat{\varphi}(g_I)|\emptyset\rangle=0.
\ee
By convention, it holds that $\langle\emptyset|\emptyset\rangle=1$. Exciting a one-particle GFT state over the Fock vacuum is expressed by
\be
|g_I\rangle=\hat{\varphi}^{\dagger}(g_I)|\emptyset\rangle
\ee
and understood as the creation of a single open $4$-valent LQG spin network vertex or of its dual tetrahedron. 

The GFT field operators obey the Canonical Commutation Relations (CCR)
\be
\bigl[\hat{\varphi}(g),\hat{\varphi}^{\dagger}(g')\bigr]=\mathbb{1}_G(g,g')~\textrm{and}~\bigl[\hat{\varphi}^{(\dagger)}(g),\hat{\varphi}^{(\dagger)}(g')\bigr]=0.
\ee
The delta distribution $\mathbb{1}_G(g,g')=\int_G dh\prod_{I}\delta(g_I h g'^{-1}_{I})$ on the space $G^d/G$ is compatible with the imposition of gauge invariance at the level of the fields as in Eq. (\ref{gaugeinvariance}).\footnote{The case of noncompact group $G$ is discussed in Ref. \cite{GFC}.} 

Using this, properly symmetrized many particle states can be constructed over the Fock space by
\be\label{manyparticlestate}
|\psi\rangle=\frac{1}{\sqrt{N!}}\sum_{P\in S_{N}}P\int(dg)^{d N} \psi(g_I^1,...,g_I^N)\prod_{i=1}^N\hat{\varphi}^{\dagger}(g^i_I)|\emptyset\rangle,
\ee
with the wave functions $\psi(g_I^1,...,g_I^N)=\langle g_I^1,...,g_I^N|\psi\rangle$. Such states correspond to the excitation of $N$ open disconnected spin network vertices. The contruction of such multiparticle states is needed for the description of extended quantum $3$-geometries.

Using this language, one can set up second-quantized Hermitian operators to encode quantum geometric observable data. In particular, an arbitrary one-body operator assumes the form
\be\label{observable}
\hat{\mathcal{O}}=\int (dg)^d\int (dg')^d~\hat{\varphi}^{\dagger}(g_I)\mathcal{O}(g_I,g_{I}')\hat{\varphi}(g_{I}'),
\ee
with $O(g_I,g_{I}')=\langle g_I|\hat{o}|g_{I}'\rangle$ given in terms of the matrix elements of the first-quantized operators $\hat{o}$. For Hermitian operators, these have to suffice of course $\mathcal{O}(g_I,g_{I}')=(\mathcal{O}(g_I',g_{I}))^{*}$. For example, the number operator is given by
\be
\hat{N}=\int (dg)^d \hat{\varphi}^{\dagger}(g_I)\hat{\varphi}(g_{I}).
\ee
Strictly speaking, $N$ exists only in the zero-interaction representation which is when all representations of the CCRs are equivalent to the Fock representation, as is well known within the context local QFTs \cite{FocknonFock}. Another relevant operator encoding geometric information is the vertex volume operator 
\be
\hat{V}=\int (dg)^d \int (dg')^d~\hat{\varphi}^{\dagger}(g_I)V(g_I,g_I')\hat{\varphi}(g_{I}'),
\ee
wherein $V(g_I,g_I')$ is given in terms of the LQG volume operator between two single-vertex spin networks and an analogous expression holds for the LQG area operator \cite{GFC,GFCReview,GFTOperators}.

\subsection{Group Field Theory Condensate Cosmology}\label{GFC}
The cosmology of the very early universe provides a natural setting in
which quantum gravity effects can be expected to have played a decisive role. The GFTCC research program attempts to describe cosmologically relevant geometries by applying the previously summarized techniques. 

In this context, the goal is to model homogeneous continuum $3$-geometries and their cosmological evolution by means of particular multiparticle GFT states, i.e. condensate states, and their effective dynamics. A possible mechanism which could lead to such condensate states, is suggested by the concept of phase transitions in GFT. As explained in the Introduction, the FRG analysis of specific GFT models has found IR fixed points in all cases investigated so far, suggesting a phase transition from a symmetric to a broken/condensate phase \cite{TGFTFRG,GFTRGReview}. The condensate corresponds to a nonperturbative vacuum of a GFT model described by a large sample of bosonic GFT quanta which all settle into a common ground state away from the Fock vacuum. To confirm the occurrence of such a transition a mean field analysis of the broken phase has to be undertaken. Ideally, the effective dynamics of the resulting effective geometry should admit a description in terms of the one given by General Relativity for the corresponding classical geometry, perhaps up to modifications \cite{GFC, GFCReview}.

\subsubsection{Condensate states}
In the following, we briefly recapitulate the motivation for why GFT condensate states serve as a good ansatz to effectively capture the physics of homogeneous continuum spacetimes following Refs. \cite{GFC,GFCReview} and review important aspects of their construction. We turn then to the extraction of the effective dynamics from the microscopic GFT action. 

In the case of spatial homogeneity, which is relevant to us, it is possible to reconstruct the geometry from any point as the metric is the same everywhere.\footnote{The procedure to reconstruct the spatial metric by means of the information encoded in the quantum state is briefly adumbrated in appendix \ref{AppendixA}.} This \textit{homogeneity criterion} translates on the level of GFT states to the requirement that all quanta occupy the same quantum geometric state. This is the reason for choosing GFT condensate states as the main ingredient for GFTCC in close analogy to the theory of real Bose condensates \cite{BECs}. Furthermore, for a state to encode in some adequate limit information allowing for the description of a smooth metric $3$-geometry, one assumes that a large constituent number $N$ will lead to a good \textit{approximation of the continuum}. Moreover, the simplicial building blocks are required to be almost flat
. This \textit{near flatness condition} translates on the level of the states to the requirement that the probability density is concentrated around small values of the curvature. Finally, for a classical cosmological spacetime to emerge from a given quantum state, it should exhibit \textit{semiclassical properties}. Crucially, condensate states automatically fulfill such a desirable feature because they are coherent states and as such exhibit, in a certain sense, ultraclassical behavior by saturating the number-phase uncertainty relation and are thus the quantum states which are the closest to classical waves. We will discuss the construction of such states and their properties in the following.

Using the Fock representation of GFT as recapitulated in appendix \ref{AppendixB}, we decompose the field operator $\hat{\varphi}(g_I)$ in terms of annihilation operators $\{\hat{c}_i\}$ of single-particle quantum geometry states $\{|i\rangle\}$ yielding
\be\label{fieldoperatorFockdecomposition}
\hat{\varphi}(g_I)=\sum_i \psi_i(g_I)\hat{c}_i.
\ee
Following the logic of the Bogoliubov approximation valid for ultracold, non- to weakly interacting and dilute Bose condensates \cite{BogoliubovAnsatz, BECs}, if the ground state $i=0$ has a macroscopic occupation, one separates this expression into a condensate term and one for all the remaining noncondensate components. This yields
\be
\hat{\varphi}(g_I)=\psi_0(g_I)c_0+\sum_{i\neq 0}\psi_i\hat{c}_i,
\ee
where one replaces the operator $\hat{c}_0$ by the c-number $c_0$ so that the average occupation number of the ground state is given by $N=\langle \hat{c}_0^{\dagger}\hat{c}_0\rangle$. In the next step one redefines $\sigma\equiv\sqrt{N}\psi_0$ as well as $\delta\hat{\varphi}\equiv\sum_{i\neq 0}\psi_i\hat{c}_i$ giving rise to
\be\label{BogoliubovAnsatz} 
\hat{\varphi}(g_I)=\sigma(g_I)+\delta\hat{\varphi}(g_I),
\ee
where $\psi_0$ is normalized to $1$. This ansatz is only justified if the ground state is macroscopically occupied, i.e., $N\gg 1$ and the fluctuations $\delta\hat{\varphi}$ are regarded as small. One calls the classical field $\sigma(g_I)$ the mean field of the condensate which assumes the role of an order parameter. Making use of the particle density $n(g_I)=|\sigma(g_I)|^2$ and a phase characterizing the coherence properties of the condensate, we write the mean field in polar form as
\be
\sigma(g_I)=\sqrt{n}~e^{i\theta(g_I)}.
\ee
This illustrates that the order parameter can always be multiplied by an arbitrary phase factor without affecting the physical measurement. This behavior is identified as a global $\mathrm{U}(1)$-symmetry of the system which is associated with the conservation of the total particle number. Upon BEC phase transition a particular phase is chosen which amounts to the spontaneous breaking of this symmetry.

By construction, the Bogoliubov ansatz (\ref{BogoliubovAnsatz}) gives rise to a nonzero expectation value of the field operator, i.e., $\langle\hat{\varphi}(g_I)\rangle\neq 0$, indicating that the condensate state is in, or rather close to, a coherent state.

Concretely, the simplest choice for the order parameter is provided by a condensate state
\be
|\sigma\rangle=A~e^{\hat{\sigma}}|\emptyset\rangle,~~~\hat{\sigma}=\int (dg)^4~\sigma(g_I)\hat{\varphi}^{\dagger}(g_I),
\ee
which is constructed from quantum tetrahedra all encoding the same discrete geometric data.\footnote{In principle, more complicated types of composite states can be considered as in Ref. \cite{GFC}.} It defines a nonpeturbative vacuum over the Fock space. The normalization factor is given by
\be
A=e^{-\frac{1}{2}\int(dg)^4~|\sigma(g)|^2}.
\ee
We require in addition to the right invariance as in Eq. (\ref{gaugeinvariance}) invariance under the left diagonal action of $G$, i.e., $\sigma(k g_I)=\sigma(g_I)$ for all $k\in G$. The latter encodes the invariance under local frame rotations.

Such states are coherent because they are eigenstates of the field operator,
\be
\hat{\varphi}(g_I)|\sigma\rangle=\sigma(g_I)|\sigma\rangle,
\ee
such that indeed $\langle\hat{\varphi}(g_I)\rangle=\sigma(g_I)\neq 0$ holds (as long as $|\sigma\rangle$ is not the Fock vacuum). Due to this property the expectation value of the number operator immediately yields the average particle number
\be
N=\int(dg)^4~|\sigma(g_I)|^2<\infty.
\ee 
It is of course only possible to use such a condensate state for the description of a macroscopic homogeneous universe, if the number of quanta is $N\gg 1$ but finite. If the number operator is well defined and its expectation value is finite, the states used here are Fock coherent states (cf. appendices \ref{AppendixB} and \ref{AppendixC}). By construction, such a description is only valid for noninteracting or weakly interacting condensates. Toward the strongly interacting regime, it has to be replaced by one given in terms of non-Fock coherent states, as the appendices \ref{AppendixB} and \ref{AppendixC} suggest.

\subsubsection{Effective dynamics}

After having discussed the construction of suitable states, let us briefly summarize how the effective condensate dynamics can be obtained from the underlying GFT dynamics as in Refs. \cite{GFC,GFCReview}.

This is done by using the infinite tower of Schwinger-Dyson equations
\be
0=\delta_{\bar{\varphi}}\langle \mathcal{O}[\varphi,\bar{\varphi}]\rangle=\biggl\langle \frac{\delta\mathcal{O}[\varphi,\bar{\varphi}]}{\delta\bar{\varphi}(g_I)}-\mathcal{O}[\varphi,\bar{\varphi}]\frac{\delta S[\varphi,\bar{\varphi}]}{\delta\bar{\varphi}(g_I)}\biggr\rangle,
\ee
where $\mathcal{O}$ is a functional of the fields. One extracts an expression for the effective dynamics by setting $\mathcal{O}$ equal to the identity. This leads to
\be
\biggl\langle \frac{\delta S[\varphi,\bar{\varphi}]}{\delta\bar{\varphi}(g_I)}\biggr\rangle=0
\ee
with the action $S[\varphi,\bar{\varphi}]$ as in Eq. (\ref{classicalaction}). When the expectation value is taken with respect to the condensate state $|\sigma\rangle$, one obtains the analog of the Gross-Pitaevskii (GP) equation for real Bose condensates
\be\label{GPGFC}
\int (dg')^4\mathcal{K}(g_I,g'_I)\sigma(g'_I)+\frac{\delta
V}{\delta\bar{\sigma}(g_I)}=0.
\ee
This is in general a nonlinear and nonlocal equation for the dynamics of the mean field $\sigma$ and is interpreted as a quantum cosmology equation. In analogy to the GP equation, it has no direct probabilistic interpretation. These features might appear as a problem when trying to relate the GFTCC framework to LQC \cite{LQC} or Wheeler-DeWitt (WdW) quantum cosmology \cite{WdWQC}. However, they do not pose a problem for the direct extraction of cosmological predictions from the full theory.  We refer to Refs.  \cite{GFC,GFCExample,GFCEmergentFriedmann,GFCEmergentFriedmann2,GFCEmergentFriedmann3}, where it has been demonstrated how a Friedmann-like evolution equation can be derived from such an effective dynamics of specific GFT condensates.

\section{Toward the mean field analysis of an interacting GFTCC model}\label{GFCInteractions}

In the following, we proceed with analyzing the quantum dynamics of a particular GFT/GFTCC model in the free and interacting cases. The larger scope of such an analysis is to see whether one can construct particular condensate solutions which admit e.g. an interpretation in terms of smooth continuous $3$-geometries and are in line with the geometrogenesis picture. 

We first review a free model and discuss how the general solution for an isotropic condensate is obtained from the equation of motion of the mean field $\sigma$ in subsection \ref{subsectionfree}. By doing so, we follow closely Ref. \cite{GFCExample} and further elaborate special solutions. We extensively discuss the geometric interpretation of such solutions by analyzing their curvature properties and by computing the expectation values of the volume and area operators imported from LQG. In subsection \ref{subsectioninteractions} we then introduce two types of combinatorially local interaction terms in subsection \ref{subsectioninteractionssetup} and firstly treat them in subsection \ref{subsectioninteractionsperturbation} as perturbations of the aforementioned free solutions. We study solutions around the nontrivial minima of the resulting effective potentials in subsection \ref{subsectioninteractionslocalminima} and discuss the expectation values of the LQG volume and area operators in this case. Finally, we conclude the analysis by interpreting the obtained results.

In order to study the quantum dynamics of the mean field $\sigma$ as in Eq. (\ref{GPGFC}), at first we have to specify the details of the action
\be
S[\varphi,\bar{\varphi}]=\int(dg)^4(dg')^4\bar{\varphi}(g_I)\mathcal{K}(g_I,g'_I)\varphi(g'_I)+\mathcal{V}[\varphi,\bar{\varphi}].
\ee
Most generally, one could study the evolution in relational terms by adding a free massless scalar field $\phi$ into the action. For the GFT field one would then have $\varphi=\varphi(g_I,\phi)$ where $\phi\in\mathbb{R}$ accounts for the relational clock as discussed in Ref. \cite{GFCEmergentFriedmann}. In these terms, the local kinetic operator is given by
\be\label{kineticoperatorgeneral}
\mathcal{K}=\delta(g'_{I}g^{-1}_{I})\delta(\phi'-\phi)\biggl[-(\tau\partial_{\phi}^2+\sum_{I=1}^4\Delta_{g_I})+m^2\biggr].
\ee
The signs of the terms appearing in $\mathcal{K}$ are chosen such that the functional $S$ in the partition function $Z$ is bounded from below. For $\tau>0$, the operator $-(\tau\partial_{\phi}^2+\sum_{I=1}^4\Delta_{g_I})$ is positive and also this choice accounts for the correct coupling of matter to gravity as noticed in Ref. \cite{GFC}. The Laplacian on the group manifold is motivated by the renormalization group analysis of GFT models where one can show that it is generated by radiative corrections (cf. \cite{GFTRGReview}). The "mass term" is related to the GFT/spin foam correspondence, as it corresponds to the spin foam edge weights.\footnote{Freezing the kinetic operator to the identity, would then lead to the ultralocal truncation of the model and establish the above-discussed correspondence between certain GFT and spin foam models \cite{GFTSF} for an appropriate choice of interaction term.} Throughout the remainder of this article, we will focus on "static" mean fields, i.e. $\sigma(g_I,\phi)=\sigma(g_I)$. The choice of action is then finalized by selecting the interaction term $\mathcal{V}$. 

\subsection{A static free model}\label{subsectionfree}

In a first approximation, we neglect all interactions and set $\mathcal{V}=0$. Using Eqs. (\ref{GPGFC}) and (\ref{kineticoperatorgeneral}), this yields
\be\label{FreeGP}
\biggl[-\sum_{I=1}^4\Delta_{g_I}+m^2\biggr]\sigma(g_I)=0.
\ee
To find solutions to this dynamical equation, we introduce coordinates on the $\textrm{SU}(2)$ group manifold, use invariance properties of $\sigma(g_I)$ and apply symmetry reductions, where we closely follow the results of Ref. \cite{GFCExample} and elaborate them where needed.

To this aim, assume that the connection in the holonomy $g=\mathcal{P}e^{i\int_e A}$ remains approximately constant along the link $e$ with length $\ell_0$ in the $x$-direction, which yields $g\approx e^{i\ell_0 A_x}$. In the polar decomposition, this gives 
\be
g=\cos(\ell_0||\vec{A}_x||)\mathbbm{1}+i\vec{\sigma}\frac{\vec{A}_x}{||\vec{A}_x||}\sin(\ell_0||\vec{A}_x||),
\ee 
with the $\mathfrak{su(2)}$-connection $A_x=\vec{A}_x\cdot\vec{\sigma}$ and the Pauli matrices $\{\sigma_i\}_{i=1...3}$. In the next step, we introduce the coordinates $(\pi_0,...,\pi_3)$ together with $\pi_0^2+...+\pi_3^2=1$ which specifies an embedding of $\mathrm{SU}(2)\cong S^3$ into $\mathbb{R}^4$. Due to the isomorphism $\textrm{SO}(3)\cong \textrm{SU}(2)/\mathbb{Z}_2$, the choice of sign in $\pi_0=\pm\sqrt{1-\vec{\pi}^2}$ corresponds to working on one hemisphere of $S^3$. With the identification
\be\label{coordinatesongroup}
\vec{\pi}=\frac{\vec{A}_x}{||\vec{A}_x||}\sin(\ell_0||\vec{A}_x||),
\ee
we can parametrize the holonomies as
\be
g(\vec{\pi})=\sqrt{1-\vec{\pi}^2}\mathbbm{1}+i\vec{\sigma}\cdot\vec{\pi},~~||\vec{\pi}||\leq 1,
\ee
where $||\vec{\pi}||=0$ corresponds to the pole of the hemisphere and $||\vec{\pi}||=1$ marks the equator. In these coordinates the Haar measure becomes
\be
dg=\frac{d\vec{\pi}}{\sqrt{1-\vec{\pi}^2}}.
\ee
Using the Lie derivative on the group manifold acting on a function $f$, one has for the Lie algebra elements
\be
\vec{B}f(g)\equiv i\frac{d}{dt}f(e^{\frac{i}{2}\vec{\sigma}t}g)|_{t=0}.
\ee
With this the Laplace-Beltrami operator $\vec{B}^2=-\Delta_{g}$ in terms of the coordinates $\vec{\pi}$ on $\mathrm{SU}(2)$ is given by
\be 
-\Delta_gf(g)=-[(\delta^{ij}-\pi^i\pi^j)\partial_i\partial_j-3\pi^i\partial_i]f(\vec{\pi}).
\ee
This applies to all group elements $g_I$, $I=1,...,4$ dressing the spin network vertex dual to the quantum tetrahedron. 

In the most general case, the left and right invariance implies that $\sigma(g_I)$ lives on the six-dimensional domain space $\textrm{SU}(2)\backslash\textrm{SU}(2)^4/\textrm{SU}(2)$. It is thus parametrized by six invariant coordinates $\pi_{IJ}=\vec{\pi}_{I}\cdot\vec{\pi}_J$, with $I,J=1,2,3$ and $0\leq|\pi_{IJ}|\leq 1$. 

Using the above, Eq. (\ref{FreeGP}) gives rise to a rather complicated partial differential equation. To find solutions, one imposes a symmetry reduction by considering functions $\sigma$ which only depend on the diagonal components $\pi_{II}$ and, furthermore, are assumed to be all equal. Together with Eq. (\ref{coordinatesongroup}), this yields
\be
p\equiv\pi_{II}=\sin^2(\ell_0||\vec{A}_x||).
\ee
Using this, one can rewrite Eq. (\ref{FreeGP}) as
\be\label{masterequation}
-\biggl[2p(1-p)\frac{d^2}{dp^2}+(3-4p)\frac{d}{dp}\biggr]\sigma(p)+\mu\sigma(p)=0,
\ee
with $\mu\equiv\frac{m^2}{12}$ and $p\in[0,1]$ for which analytic solutions can be found \cite{GFCExample}.\footnote{Observe that this symmetry reduction should not be confused with those performed in WdW quantum cosmology or LQC, since it is applied after quantization onto the quantum state and not before.}

Indeed, it makes sense to refer to this symmetry reduction (to just one variable $p$) as an isotropization. Retrospectively, this can be seen when rewriting Eq. (\ref{masterequation}) using $p\equiv \sin^2(\psi)$. With this we obtain
\be\label{masterequationangle}
-[\frac{d^2}{d\psi^2}+2\cot(\psi)\frac{d}{d\psi}]\sigma(\psi)+2\mu\sigma(\psi)=0,~~~\psi\in[0,\pi/2],
\ee
which can be compared to the Laplacian on a hemisphere of $S^3$ acting on a function $\sigma(\phi,\theta,\psi)$, given by
\be\label{LaplacianHemisphere}
-\Delta \sigma(\phi,\theta,\psi)=-\frac{1}{\sin^2(\psi)}\biggl[\frac{\partial}{\partial\psi}(\sin^2(\psi)\frac{\partial}{\partial\psi}\sigma)+\Delta_{S^2}\sigma\biggr],
\ee
with $\phi\in[0,2\pi]$, $\theta\in[0,\pi]$ and $\psi\in[0,\pi/2]$. The function $\sigma$ is called isotropic or zonal if it is independent of $\phi$ and $\theta$ \cite{AnalysisonLieGroups}. These are spherically symmetric eigenfunctions of $-\Delta_{S^2}$ for which Eq. (\ref{masterequationangle}) is equal to Eq. (\ref{LaplacianHemisphere}). Hence, the symmetry reduction can be seen as explicitly restricting the rather general class of condensates to a representative with a clearer geometric interpretation.\newline
The general solution to Eq. (\ref{masterequation}) is given by
\begin{multline}\label{generalsolution}
\sigma(p)=\sqrt[4]{\frac{1-p}{p}}[a~P^{\frac{1}{2}}_{\frac{1}{2}\sqrt{1-2\mu}-1}(2p-1)~+~\\ b~Q^{\frac{1}{2}}_{\frac{1}{2}(\sqrt{1-2\mu}-1)}(2p-1)],
\end{multline}
with $a,b\in\mathbb{C}$ and $P,Q$ are associated Legendre functions of the first and second kinds, respectively. With respect to the measure induced from the full Fock space, one yields for the average particle number
\be
\begin{split}
N&=\int (dg)^3 |\sigma(g_1,g_2,g_3)|^2\\ &= 2\pi\int dp \sqrt{\frac{p}{1-p}}|\sigma(p)|^2<\infty.
\end{split}
\ee

In the following, we want to specify the possible values of $\mu$ in the symmetry reduced case by means of discussing the spectrum of the operator $-\sum_I\Delta_{g_I}$. Its self-adjointness and positivity imply that its eigenvalues $\{m^2\}$ lie in $\mathbb{R}_0^{+}$. The compactness of the domain space $\textrm{SU}(2)\backslash\textrm{SU}(2)^4/\textrm{SU}(2)$ entails that the spectrum is discrete and the respective eigenspaces are finite-dimensional. This also holds for the symmetry reduced case. 

To finally concretize the spectrum, we have to introduce boundary conditions, which we infer from physical assumptions. For this we can exploit that we are looking for solutions to the equation of motion which admit an interpretation in terms of smooth metric $3$-geometries and thus obey the above-mentioned near flatness condition. In the group representation, this condition concretely translates into demanding that the character of the group elements decorating the quantum tetrahedra are close to $\chi(\mathds{1}_{J_{i}})=2J_i+1$ according to Refs. \cite{GFC, GFCReview}. On the level of the mean field this leads to the requirement that the probability density is concentrated around small values of the connection or its curvature. In the symmetry reduced case this condition holds for $\sigma(p)$ if the probability density $|\sigma(p)|^2$ is concentrated around small values of the variable $p$ and tends to zero at the equator traced out at $p=1$. The latter translates into a Dirichlet boundary condition on the equator,
\be
\sigma(p)|_{p=1}=0,
\ee
which is only obeyed by the $Q$-branch of the general solution Eq. (\ref{generalsolution}). Using this, the spectrum of the Dirichlet Laplacian is given by $\mu=-2n(n+1)$ with $n\in(2\mathbb{N}_0 + 1)/2$.\footnote{The only eigenfunction of the Dirichlet Laplacian to the eigenvalue $\mu=0$ is the trivial function. In the mean field analysis of phase transitions the mean field is supposed to vanish if the driving parameter $\mu$ turns to $0$. Here, consistency with the flatness condition implies the vanishing of the mean field $\sigma$ for $\mu=0$.}\footnote{Due to the linear character of the free problem, the solutions have a rescaling invariance with respect to the chosen boundary conditions. This means that two solutions for different boundary conditions $\sigma'(1)$ can be rescaled into one another according to
\be\label{rescalinginvariance}
\frac{N[\sigma_1'(1)]}{N[\sigma_2'(1)]}=\frac{|\sigma_1'(1)|^2}{|\sigma_2'(1)|^2},
\ee
which obscures the interpretation of the quantity $N$ and other observables in the free case. This rescaling property is lost once (strong) nonlinear interactions are considered as in the next subsection.}

Equivalently, these solutions correspond to the eigensolutions of Eq. (\ref{masterequationangle}) obeying the boundary condition $\sigma(\frac{\pi}{2})=0$. They are given by
\be\label{solutionsangle}
\sigma_j(\psi)=\frac{\sin((2j+1)\psi)}{\sin(\psi)},~~~\psi\in[0,\frac{\pi}{2}]
\ee
with $j\in \frac{2\mathbb{N}_0+1}{2}$ corresponding to the eigenvalues $\mu=-2j(j+1)$. On the interval $[0,\frac{\pi}{2}]$ these solutions are exactly equal to those hyperspherically symmetric eigenfunctions of the Laplacian on $S^3$ which vanish on the equator. Furthermore, observe that these are just the characters $\chi_j(\psi)$ of the respective representation for $j$.\newline

\begin{figure}[ht]
	\centering
  \includegraphics[width=0.4\textwidth]{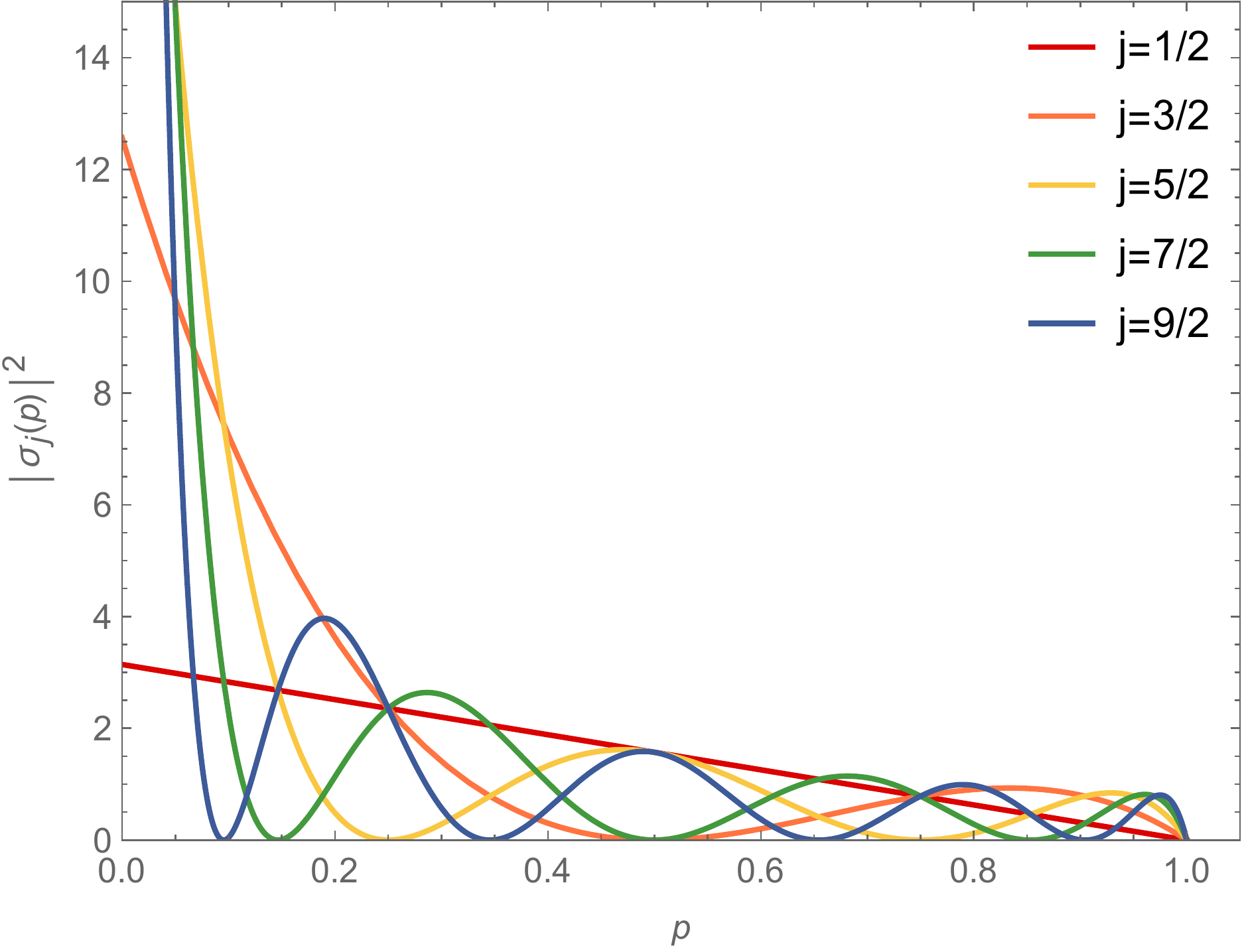}
	\caption{Probability density of the free mean field over $p$.}
	\label{fig1}
\end{figure}

\begin{figure}[ht]
	\centering
  \includegraphics[width=0.4\textwidth]{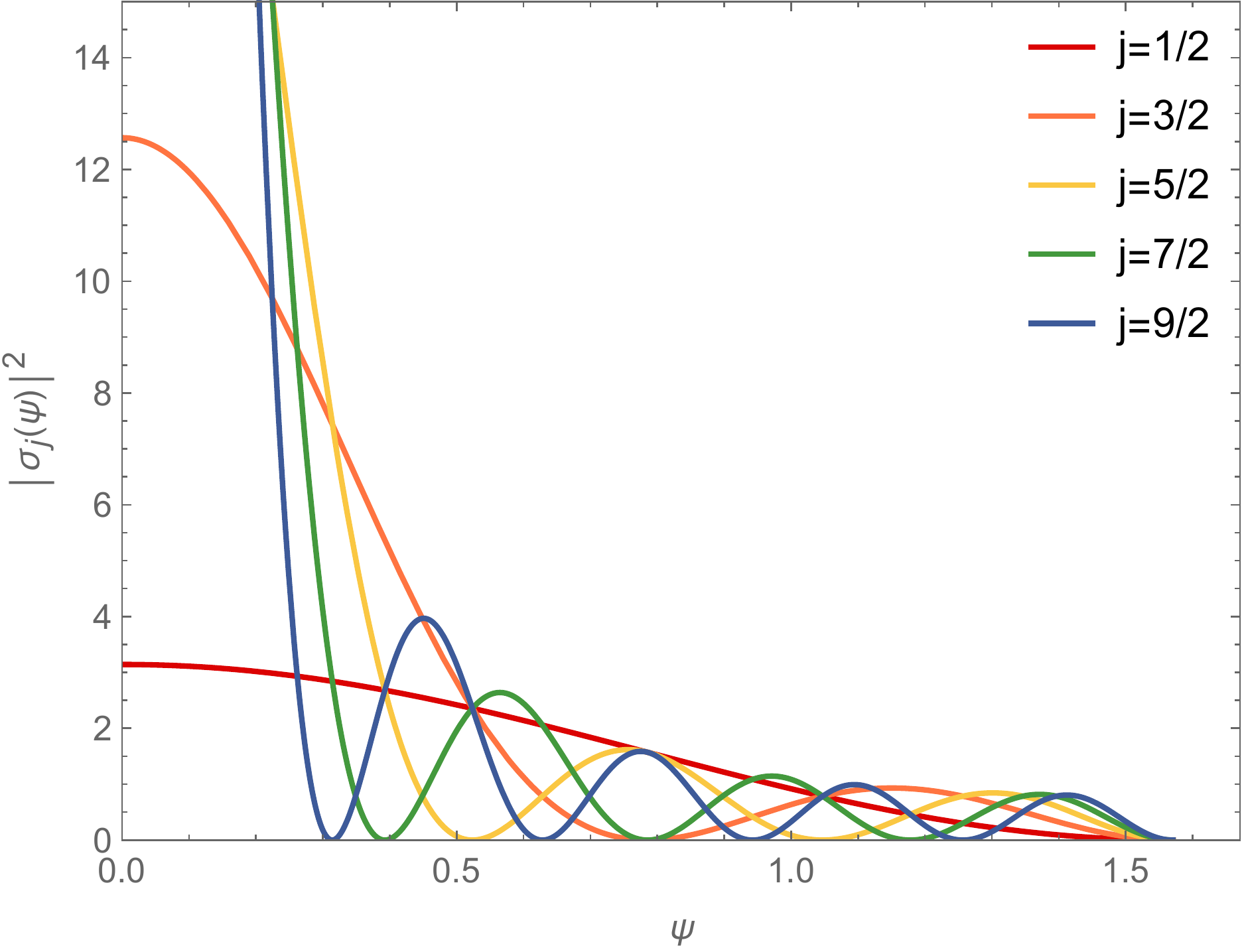}
	\caption{Probability density of the free mean field over $\psi$.}
	\label{fig2}
\end{figure}

In view of the geometric interpretation of these solutions, we want to illustrate and then discuss the behavior of the first few eigensolutions by plotting their probability density $|\sigma(p)|^2$ in Fig. \ref{fig1} or $|\sigma(\psi)|^2$ in Fig. \ref{fig2}, respectively. The plot illustrates that the probability density is concentrated around small values of the variable $p$ or $\psi$, respectively. In general, eigensolutions remain finitely peaked around $p=0$ or $\psi=0$. Solutions for slightly perturbed eigenvalues $\mu$ are infinitely peaked as $\lim_{p\to 0}|\sigma(p)|^2\sim 1/p$. 

\begin{figure}[ht]
	\centering
  \includegraphics[width=0.4\textwidth]{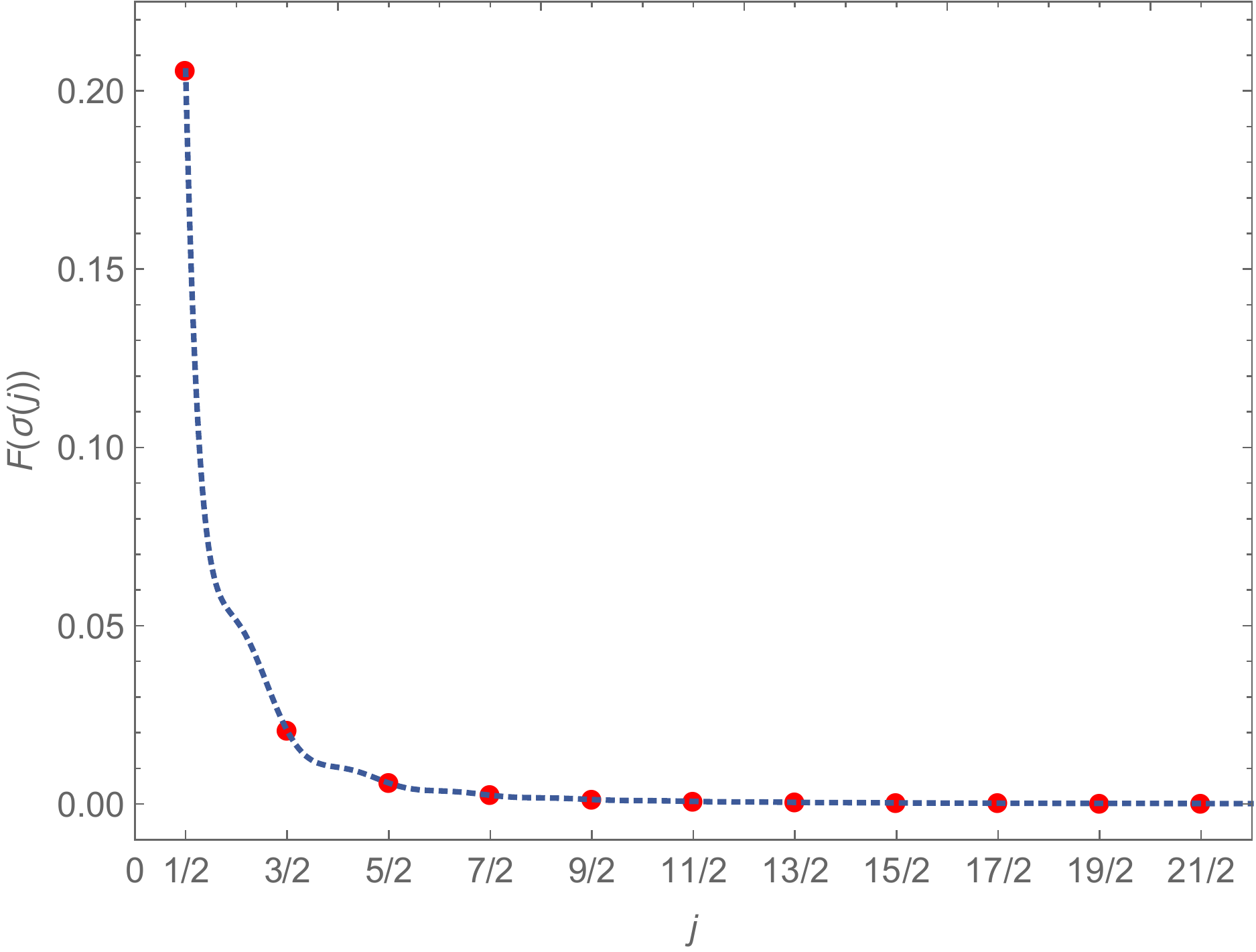}
	\caption{Un-normalized spectrum of the field strength with respect to the eigensolutions $\sigma_j(\psi)$ in arbitrary units.}
	\label{fieldstrengthfigure}
\end{figure}

A concentration of the probability density around small $p$ corresponds to a concentration around small curvature values. This is because small $p$, itself directly proportional to the gravitational connection $A$, implies small field strength via $F=D_A A$. Naively, in turn this leads to a small $3$-curvature $R$, as is known from the first-order formalism for gravity. This is important for consistency matters, meaning that the building blocks of the geometry are indeed almost flat which is needed to approximate a smooth continuum $3$-space. Around $p=1$ or $\psi=\frac{\pi}{2}$, tracing out the equator of $S^3$, the solutions vanish. The occurrence of the finite number of oscillatory maxima does not a priori pose a problem to the fulfillment of the near flatness condition since the eigensolutions are indeed concentrated around small values of $p$ or angles $\psi$, far away from the equator. For the characters of the corresponding representations, the near flatness condition means that they should be close to $\chi(\mathds{1}_{j})=2j+1$ \cite{GFC,GFCReview}. Our solutions obey this requirement since in Eq. (\ref{solutionsangle}) $\lim_{\psi\to 0}\sigma_j(\psi)$ exactly yields $2j+1$. In this light, using the solutions $\sigma_j(\psi)$, we can compute the average of the field strength\footnote{Relating $p$ to the field strength is justified when considering a plaquette $\Box$ in a face of a tetrahedron so that we can make use of the well known expression
\be\nonumber
F^k_{ab}(A)=\frac{1}{\mathrm{Tr}(\tau^k\tau^k)}\lim_{\textrm{Area}_{\Box}\to 0}\mathrm{Tr}_j\biggl(\tau^k\frac{\mathrm{hol}_{\Box_{ij}}(A)-1}{\textrm{Area}_{\Box}}\biggr)\delta_a^i\delta^j_b,
\ee
where $a,b\in\{1,2\}$ and for the $\mathfrak{su}(2)$-algebra elements $\tau_k=-\frac{i}{2}\sigma_k$ the relation $\mathrm{Tr}(\tau^k\tau^k)=-\frac{1}{3}j(j+1)(2j+1)$ holds \cite{LQC}. This yields $F^k\sim \sin^2(\psi)=p$.} $F^i\sim p$ given by
\be
\frac{\langle \hat{F}^i\rangle}{N}\sim\int_{0}^{\pi/2}d\psi\sin^2(\psi)~|\sigma_j(\psi)|^2~\hat{F}^i~/~N>0,
\ee
which is illustrated in Fig. \ref{fieldstrengthfigure}. The dots indicate the discrete contributions to the field strength for a particular $j$-mean field and show a dominance of the $1/2$-eigensolution over the others on which we comment below.
In light of the previous discussion, it may seem a bit surprising that the expectation value of the field strength is nonzero despite the fact that $p=0$ is the most probable value of the corresponding mean field. However, the extended tail of the probability density with the finite oscillatory maxima accounts for the average being bigger than the most probable value. The finite value indicates that the space described by the condensate is of finite size. We will come back to this point at the end of this section.\footnote{The last word on the flatness behavior of such solutions also in the interacting case, however, lies with the analysis (of the expectation value) of a currently lacking GFT-curvature operator, as already noticed in Ref. \cite{GFCReview}.}

\begin{figure}[ht]
	\centering
  \includegraphics[width=0.4\textwidth]{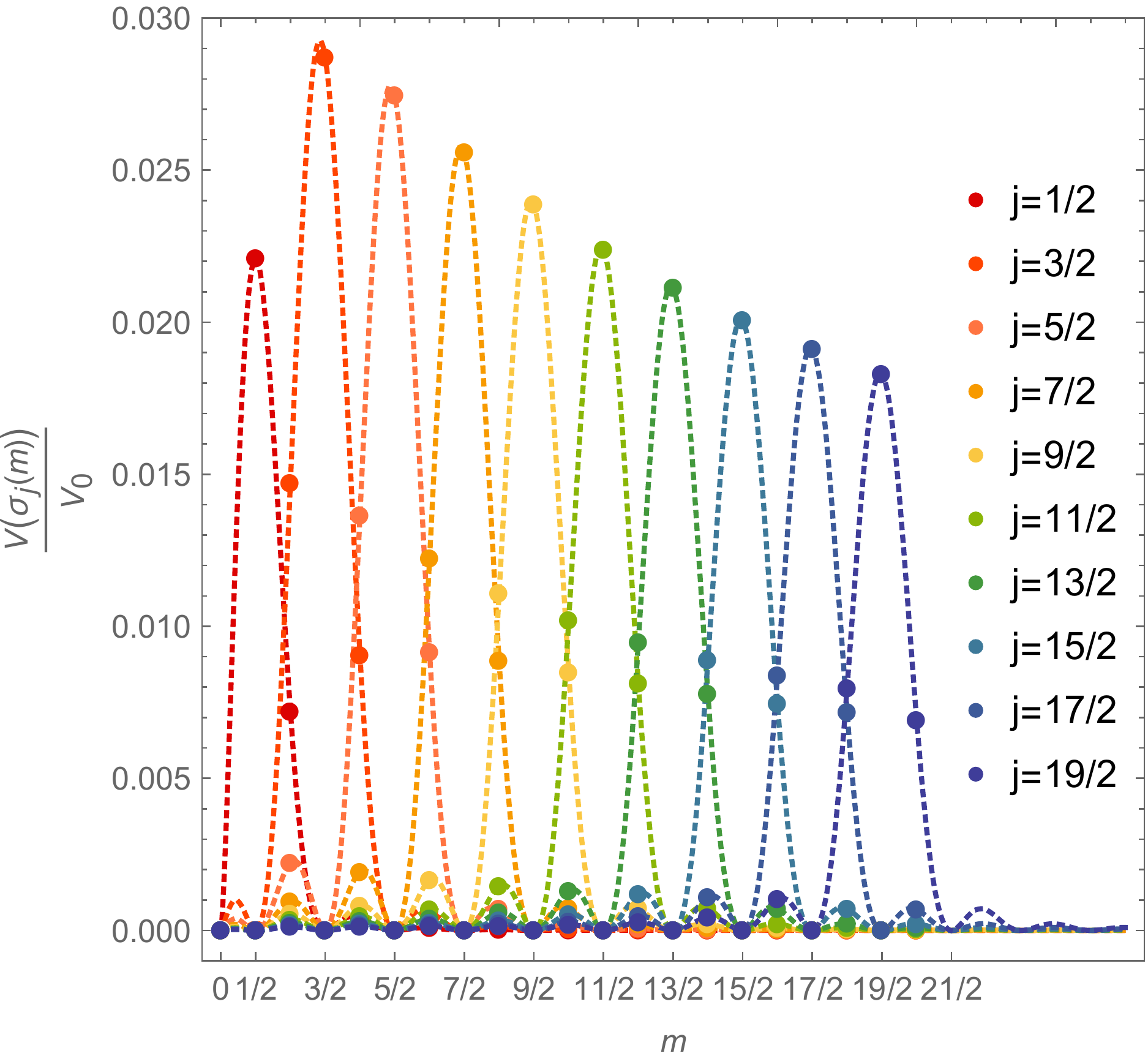}
	\caption{Normalized spectrum of the volume operator with respect to the eigensolutions $\sigma_j(\psi)$ in arbitrary units.}
	\label{fig4}
\end{figure}

\begin{figure}[ht]
	\centering
  \includegraphics[width=0.35\textwidth]{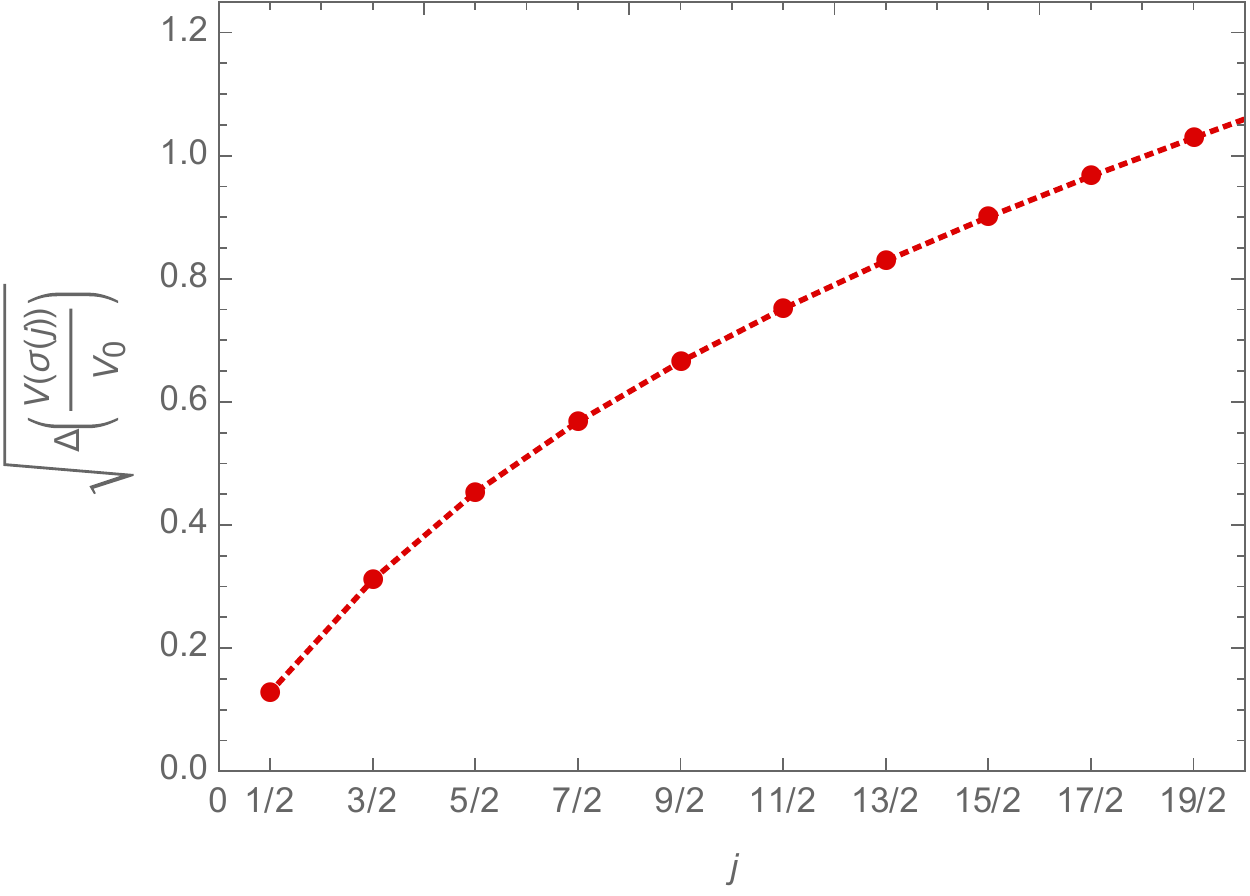}
	\caption{Standard deviation of the volume operator over $j$.}
	\label{fig5}
\end{figure}

In the last step, we want to transform our nearly flat solutions to the spin-representation which facilitates most directly the extraction of information about the LQG volume and area operators and is crucial for the geometric interpretation of the solutions. To this aim, notice that due to the left- and right-invariance of $\sigma(g_I)$, the mean field is in particular a central function on the domain space, i.e., $\sigma(h g_I h^{-1})=\sigma(g_I)$ for all $h\in \mathrm{SU}(2)$. This holds for the isotropic function $\sigma(p)$ or $\sigma(\psi)$, analogously. In this case isotropy coincides with the notion of centrality. Using the Fourier series of a central function on $\mathrm{SU}(2)$ \cite{AnalysisonLieGroups}, the Fourier series for the mean field in the angle parametrization is given by
\be 
\sigma_j(\psi)=\sum_{m\in\mathbb{N}_0/2}(2m+1)~\chi_m(\psi)~\sigma_{j;m},
\ee
with the "plane waves" given by the characters $\chi_m(\psi)=\frac{\sin((2m+1)\psi)}{\sin(\psi)}$. The Fourier coefficients are then obtained via
\be
\sigma_{j;m}=\frac{2}{\pi}\frac{1}{2m+1}\int_{0}^{\pi/2}d\psi \sin^2(\psi)~\chi_m(\psi)~\sigma_j(\psi),
\ee
and $m\in\frac{\mathbb{N}_0}{2}$. Using this, the Fourier coefficients of the solutions $\sigma_j(\psi)$ (cf. Eq. (\ref{solutionsangle})) yield
\be
\sigma_{j;m}=\frac{2}{\pi}\frac{1}{2m+1}\frac{(-1)^{\frac{2j-1}{2}}~(2j+1)~\cos\bigl(\frac{2m\pi}{2}\bigr)}{(2m-2j)~(2m+2j+2)},
\ee
with $j\in \frac{2\mathbb{N}_0+1}{2}$. In the spin-representation, the expectation value of the volume operator with respect to the mean field is decomposed as
\be
\langle \hat{V}\rangle\equiv V=V_0\sum_{m\in\mathbb{N}_0/2} |\sigma_{j;m}|^2 V_m~~\textrm{with}~~V_m\sim m^{3/2}
\ee
and $V_0\sim\ell_p^3$.\footnote{As a side remark, notice that if $\mu>0$, naively $j$ would be complex and thus also the spectra of the geometric operators such as the volume $\hat{V}$.} The normalized volume $V/V_0$ is shown in Fig. \ref{fig4} for different values of $j$.
The dots indicate the discrete contributions to the volume for a particular $j$. Eigensolutions for smaller $j$ or $|\mu|$ have a bigger volume in comparison to those with larger $j$, especially the $j=3/2$ eigensolution has the relatively biggest volume. Importantly, the volume is finite for all $j$ indicating that the space which the condensate approximates must be of finite size. Hence, a general solution which can be decomposed in terms of eigensolutions, describes a finitely sized space of which the largest contributions arise from low spin modes. Finally, Fig. \ref{fig5} illustrates the uncertainty of the volume operator, which is monotonously increasing in $j$ and indicates that its expectation value assumes a sharper value if the condensate resides in lower $j$-modes.

\begin{figure}[ht]
	\centering
  \includegraphics[width=0.4\textwidth]{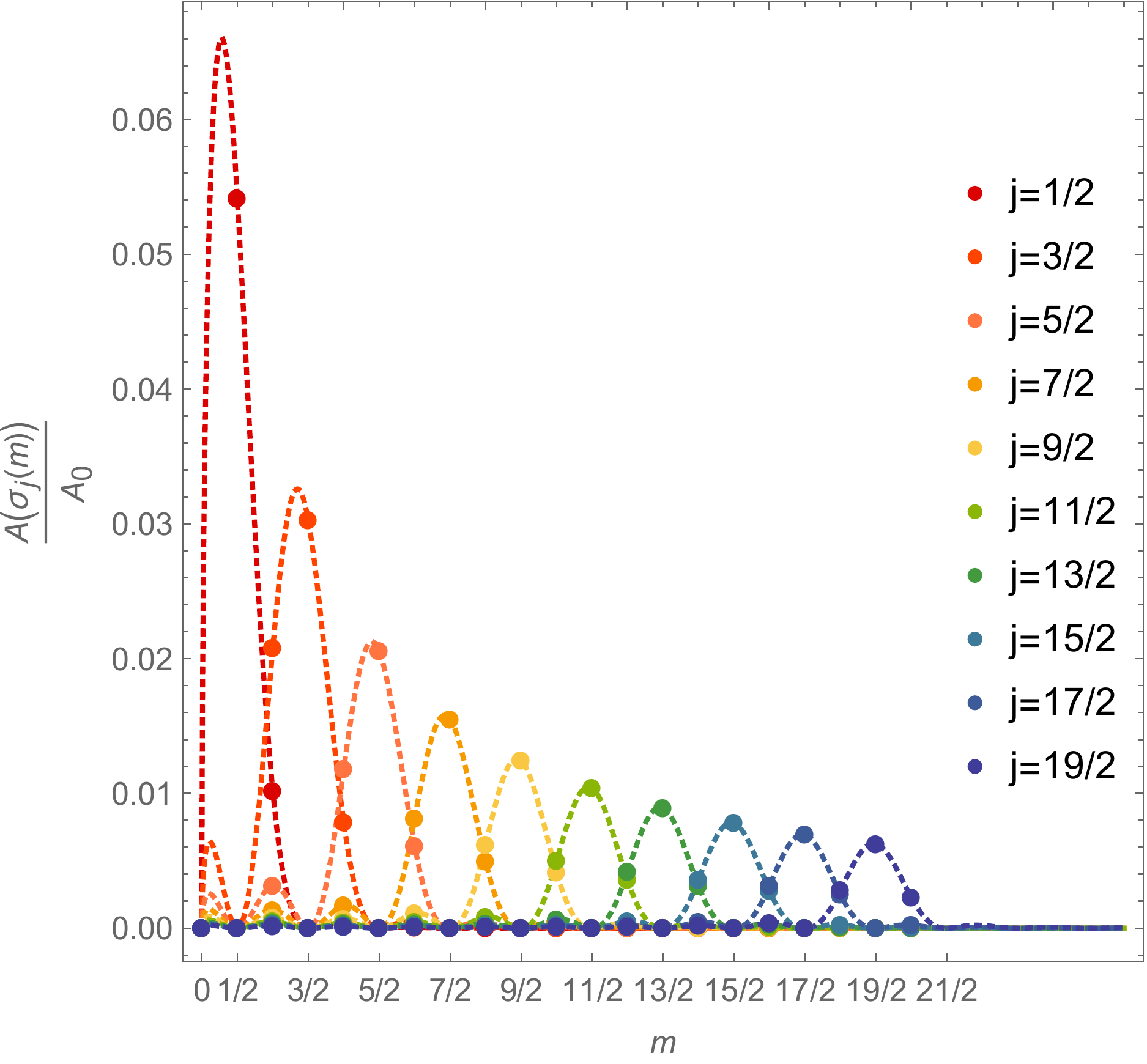}
	\caption{Normalized spectrum of the area operator with respect to the eigensolutions $\sigma_j(\psi)$ in arbitrary units.}
	\label{spectrumarea}
\end{figure}

Analogously, the expectation value of the area operator for an individual face of a quantum tetrahedron in the condensate is given as
\be
\langle \hat{A}\rangle\equiv A=A_0\sum_{m\in\mathbb{N}_0/2} |\sigma_{j;m}|^2 A_m
\ee
with $A_m\sim (m(m+1))^{1/2}$ and $A_0\sim\ell_p^2$. Depending on the solution $\sigma_j$, the spectrum of the normalized area $A/A_0$ is illustrated in Fig. \ref{spectrumarea} showing a dominance of the $1/2$-representation and otherwise with a similar interpretation as in the case of the volume operator.

From the above, it is not clear whether a certain eigensolution could be dynamically preferred over others. The near flatness condition seems to be better fulfilled by lower eigenmodes, that means for those solutions with a lower number of oscillatory maxima. These are the solutions which are mostly concentrated around small connection or curvature values. This might be connected to the recent findings of a dynamically reached low spin phase in a similar GFT condensate cosmology model \cite{GFClowspin}. In this light, it is striking that the computation of the expectation values of the volume, area and the field strength operators all display the dominance of low $j$-modes. This can be seen to be in favor of the condensate picture where the field quanta tend to condense into the same simple quantum geometric state. Below we explore the case of interacting models which is pivotal for the geometric interpretation of the solutions and the extraction of phenomenology.

We want to make a final remark about restricting our attention solely to the those solutions obeying the near flatness condition. Of course one could consider more general solutions to Eq. (\ref{masterequation}) which are not necessarily peaked around $p=0$ as in Ref. \cite{GFCExample}. Despite the fact, that such solutions cannot be interpreted in terms of smooth continuous $3$-geometries according to the near flatness condition proposed in Ref. \cite{GFC}, their properties could nevertheless be studied in a similar manner which will be done elsewhere.

\subsection{Static interacting models}\label{subsectioninteractions}

In this subsection we add to the above considered free model two different combinatorially local interaction terms, i.e. pseudopotentials, and analyze their effect on the behavior of the solutions and the expectation values of releavant operators. This enables us to study aspects of the resulting effective quantum geometries. 

One might speculate that such simplified interactions between the condensate constituents are only relevant in a continuum and large scale limit, where the true combinatorial nonlocality of the fundamental theory could be effectively hidden. This idea can perhaps be motivated by speculating that while the occurrence of UV fixed points in tensorial GFTs is deeply rooted in their combinatorial nonlocality (cf. \cite{TGFTFRG, TMRG}), the occurrence of IR fixed points, akin to Wilson-Fisher fixed points in the corresponding local QFTs, seems to be unaffected by this feature. Ultimately, rigorous RG arguments will have the decisive word whether combinatorially local interaction terms may be derived from the fundamental theory. In this way, studying the effect of pseudopotentials and trying to extract physics from the solutions can be useful to clarify the map between the microscopic and effective macroscopic dynamics of the theory and is instructive to gain experience for the treatment of the corresponding nonlocal terms which have a clearer discrete geometric interpretation. In this light, we will consider two classes of local interactions, mimicking the so-called tensorial and the above-introduced simplicial interactions.

\subsubsection{General setup of the interacting GFTCC models}\label{subsectioninteractionssetup}

The models on which we built our analysis assume an action of the form
\be\label{actionrealfieldgeneral}
S[\varphi,\bar{\varphi}]=\int(dg)^4(dg')^4\bar{\varphi}(g_I)\mathcal{K}(g_I,g'_I)\varphi(g'_I)+\mathcal{V}[\varphi,\bar{\varphi}],
\ee
with the kinetic operator
\be
\mathcal{K}=\delta(g'_{I}g^{-1}_{I})\biggl[-\sum_{I=1}^4\Delta_{g_I}+m^2\biggr]
\ee
and the general pseudopotential mimicking so-called tensorial interactions
\be
\mathcal{V}_T[\varphi]=\sum_{n\geq 2}\frac{\kappa_{n}}{n}\int (dg)^4(|\varphi(g_I)|^2)^n
\ee
which is even powered in the modulus of the field. One obtains for the equation of motion of the mean field 
\begin{equation}
\biggl[-\sum_{I=1}^4\Delta_{g_I}+m^2\biggr]\sigma(g_I)+\sigma(g_I)\sum_{n=2}\kappa_{n}(|\sigma(g_I)|^2)^{n-1}=0.
\end{equation}
Observe that the combinatorial locality implies that we do not make use of any nontrivial pairing pattern for the fields and when applying the same symmetry assumptions as above one has $\sigma(g_1,g_2,g_3,g_4)=\sigma(g,g,g,g)=\sigma(p)$. Considering only one summand for the interaction, we yield
\begin{multline}
-[2p(1-p)\frac{d^2}{dp^2}+(3-4p)\frac{d}{dp}]\sigma(p)+\mu\sigma(p)~+~\\ \kappa\sigma(p)(|\sigma(p)|^2)^{n-1}=0,
\end{multline}
with $n=2,~3,~4,...$. 

\begin{figure}[ht]
	\centering
  \includegraphics[width=0.3\textwidth]{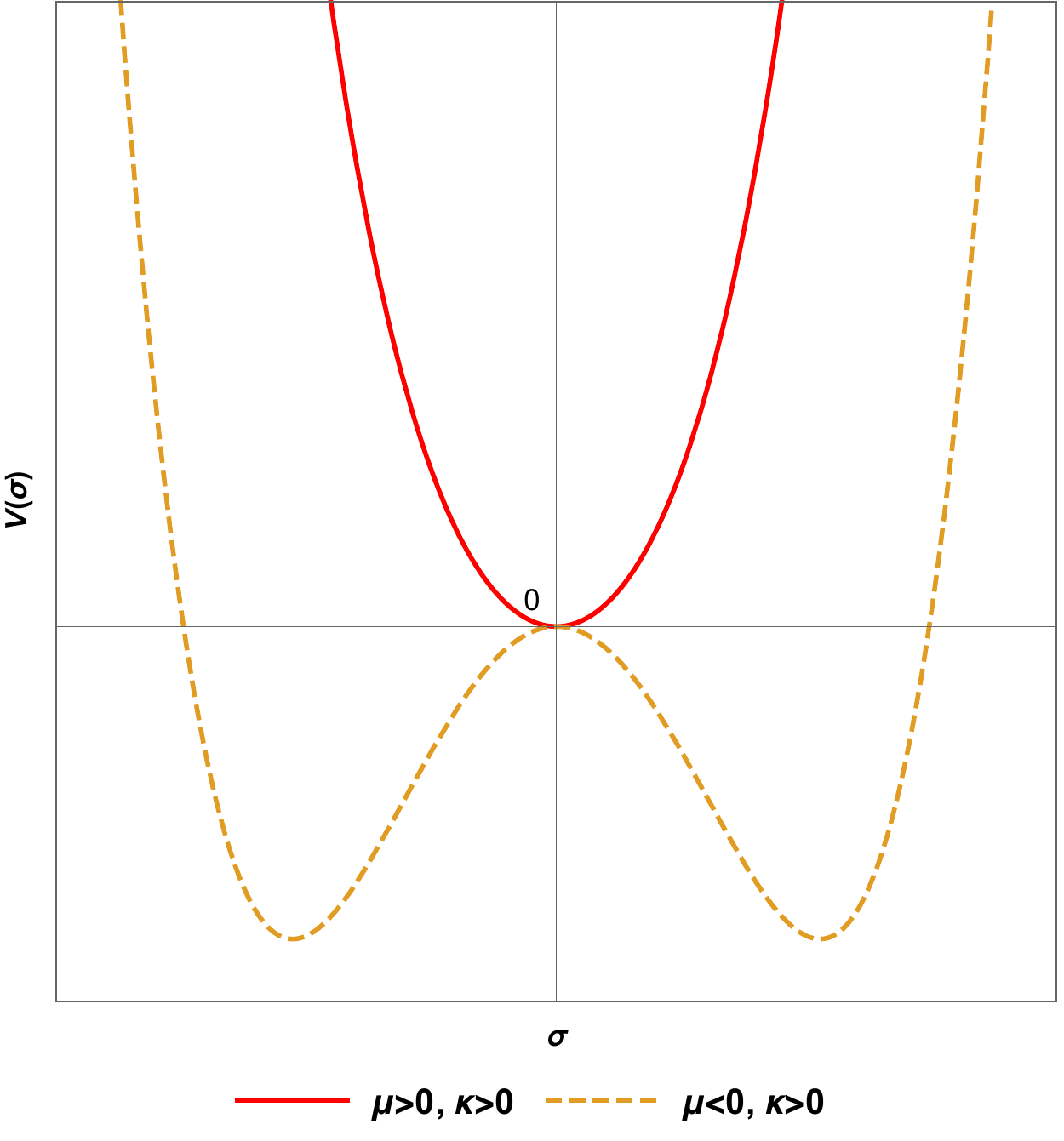}
	\caption{Plot of the effective potential $V_{eff}[\sigma]=\frac{\mu}{2}\sigma^2+\frac{\kappa}{4}\sigma^4$.}
	\label{reellesfeldfi4}
\end{figure}

In the following, we focus on the case of real-valued GFT fields and set $n=2$, for which the equation of motion reads
\be\label{odesigma4}
-[2p(1-p)\frac{d^2}{dp^2}+(3-4p)\frac{d}{dp}]\sigma(p)+\mu\sigma(p)+\kappa\sigma(p)^{3}=0,
\ee
with the effective potential
\be\label{phi4potential}
V_{eff}[\sigma]=\frac{\mu}{2}\sigma^2+\frac{\kappa}{4}\sigma^4.
\ee
The signs of the coupling constants determine the structure of the ground state of the theory. For appropriately chosen signs of $\mu$ and $\kappa$ the potential, and thus the spectrum of the theory, is bounded from below. However, only for $\mu<0$ and $\kappa>0$ one can have a nontrivial (nonperturbative) vacuum with 
\be
\langle \sigma\rangle \neq 0,
\ee
which is needed to be in agreement with the condensate state ansatz. The two distinct minima of the potential are located at $\langle\sigma_0\rangle=\pm\sqrt{-\mu/\kappa}$ where the potential has strength $V_0=-\mu^2/4\kappa$, which is lower than the value for the excited configuration $\sigma=0$. The system would thus settle into one of the minima as its equilibrium configuration and could be used to describe a condensate.\footnote{A sign change of the driving parameter $\mu$ from positive to negative values induces a spontaneous symmetry breaking of the global $\mathbb{Z}_2$-symmetry of the action specified in Eq. (\ref{actionrealfieldgeneral}). This symmetry would have guaranteed the conservation of oddness or evenness of the number of GFT quanta as it corresponds to the conserved discrete quantity $(-1)^N$. For complex-valued GFT fields the analoguous situation would correspond to the spontaneous breaking of the global $\mathrm{U}(1)$-symmetry of the action which would have guaranteed the conservation of the particle number $N$.} This potential is illustrated in Fig. \ref{reellesfeldfi4} and contrasted to the case where $\mu>0$ for which the potential is a convex function of $\sigma$ with minimum at $\langle \sigma\rangle=0$. The latter setting cannot be used to describe a condensate with $N\neq 0$. For other choices of signs, the equilibrium configuration $\langle\sigma\rangle=0$ is unstable or metastable and should be dismissed. The upshot of this discussion is that if the effective action is to represent a stable system and a condensate of GFT quanta, one must choose the signs of the coupling constants accordingly.\footnote{For real BECs, $\kappa<0$ gives an attractive interaction and only a large enough kinetic term can prevent the condensate from collapsing. In the opposite case where $\kappa>0$, the interaction is repulsive and if it dominates over the kinetic term the condensate is well described in terms of the so-called Thomas-Fermi approximation \cite{BECs}.}

Similarly, when considering the local pseudopotential mimicking the above-introduced simplicial interaction for real-valued GFT fields
\be
\mathcal{V}_S[\varphi]=\frac{\kappa}{5}\int (dg)^4\varphi(g_I)^{5},
\ee
one has
\be\label{odesigma5}
-[2p(1-p)\frac{d^2}{dp^2}+(3-4p)\frac{d}{dp}]\sigma(p)+\mu\sigma(p)+\kappa\sigma(p)^{4}=0.
\ee
For such a model the effective potential reads 
\be\label{phi5potential}
V_{eff}[\sigma]=\frac{\mu}{2}\sigma^2+\frac{\kappa}{5}\sigma^5.
\ee

\begin{figure}[ht]
	\centering
   \includegraphics[width=0.3\textwidth]{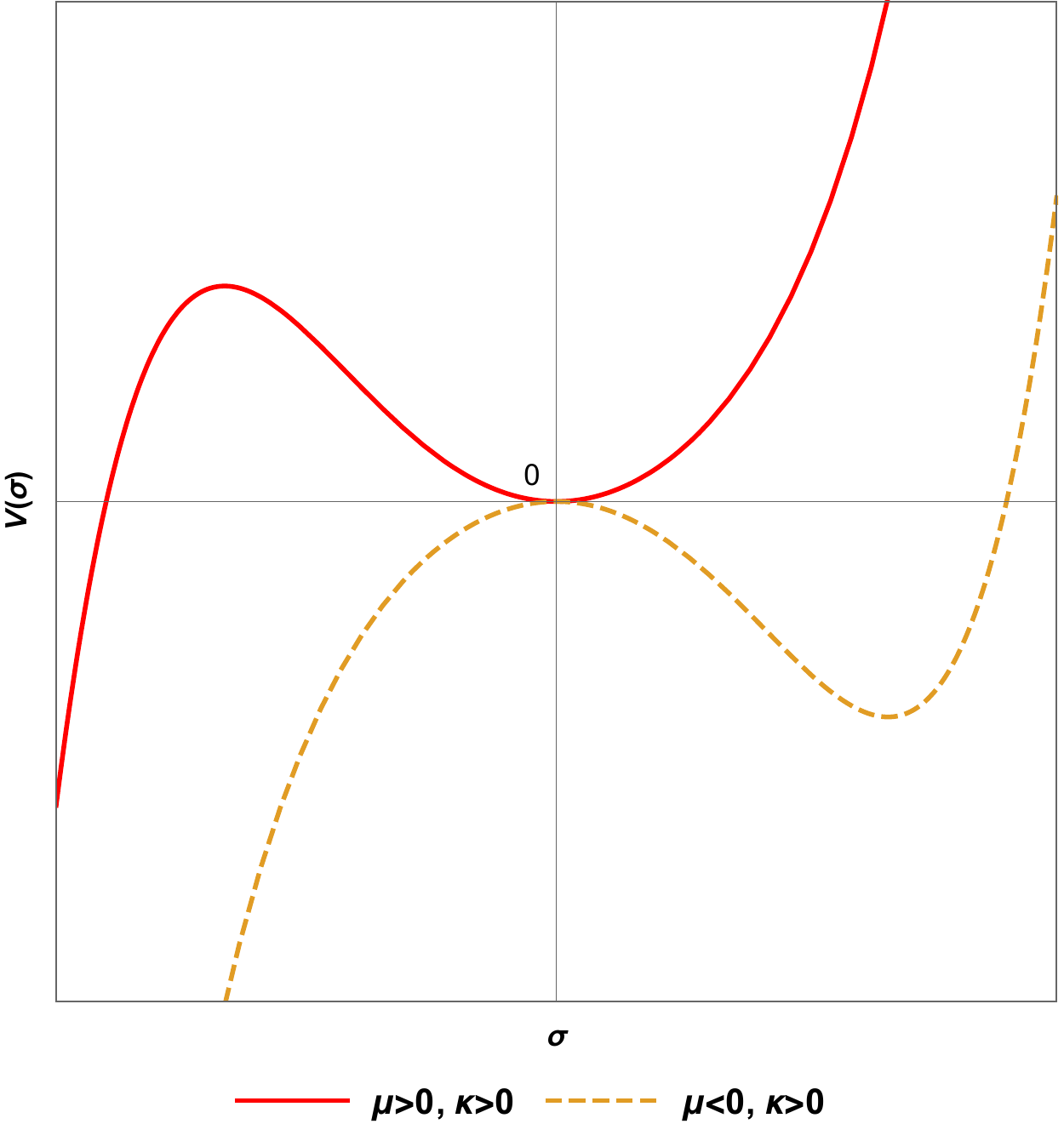}
	\caption{Plot of the effective potential $V_{eff}[\sigma]=\frac{\mu}{2}\sigma^2+\frac{\kappa}{5}\sigma^5$.}
	\label{reellesfeldfi5}
\end{figure}

We will ignore here that this potential is unbounded from below to one side.\footnote{Notice that when using four arguments in the group field $\varphi$, higher simplicial interaction terms known to be e.g. of power $16$ or $500$ would lead in the local point of view, adopted here, to bounded effective potentials $V_{eff}$ like (\ref{phi4potential}) and the discussion of their effects would be rather analogous.} Only for $(\mu<0,\kappa>0)$ or $(\mu<0,\kappa<0)$ one can have a nontrivial (nonperturbative) vacuum in agreement with the condensate state ansatz and the discussion of the choice of signs is similar to the firstly considered potential. Classically, the corresponding minima of the potential are then located at $\sigma_0=\pm\sqrt[3]{\mp\mu/\kappa}$ where the potential has strength $V_0=(\mp\mu/\kappa)^{2/3}(3\mu/10)$. This is illustrated in Fig. \ref{reellesfeldfi5} for one case and contrasted to the situation where $\mu>0$ which would lead to $\langle \sigma\rangle=0$.

\subsubsection{Perturbation of the free case}\label{subsectioninteractionsperturbation}

In a first step, we consider the interaction term as a perturbation of the free case discussed in subsection \ref{subsectionfree} using the same boundary conditions $\sigma(1)=0$ and different $\sigma'(1)$ to solve numerically the nonlinear differential equations  (\ref{odesigma4}) and (\ref{odesigma5}), respectively. By following closely the procedure adopted in the free case, we compute the effect of perturbations onto the probability densities and the spectra of geometric operators. In this way we obtain a clear qualitative picture of the effect of interactions by comparing the results to the ones obtained for the free case. 

\begin{figure}[ht]
	\centering
   \includegraphics[width=0.4\textwidth]{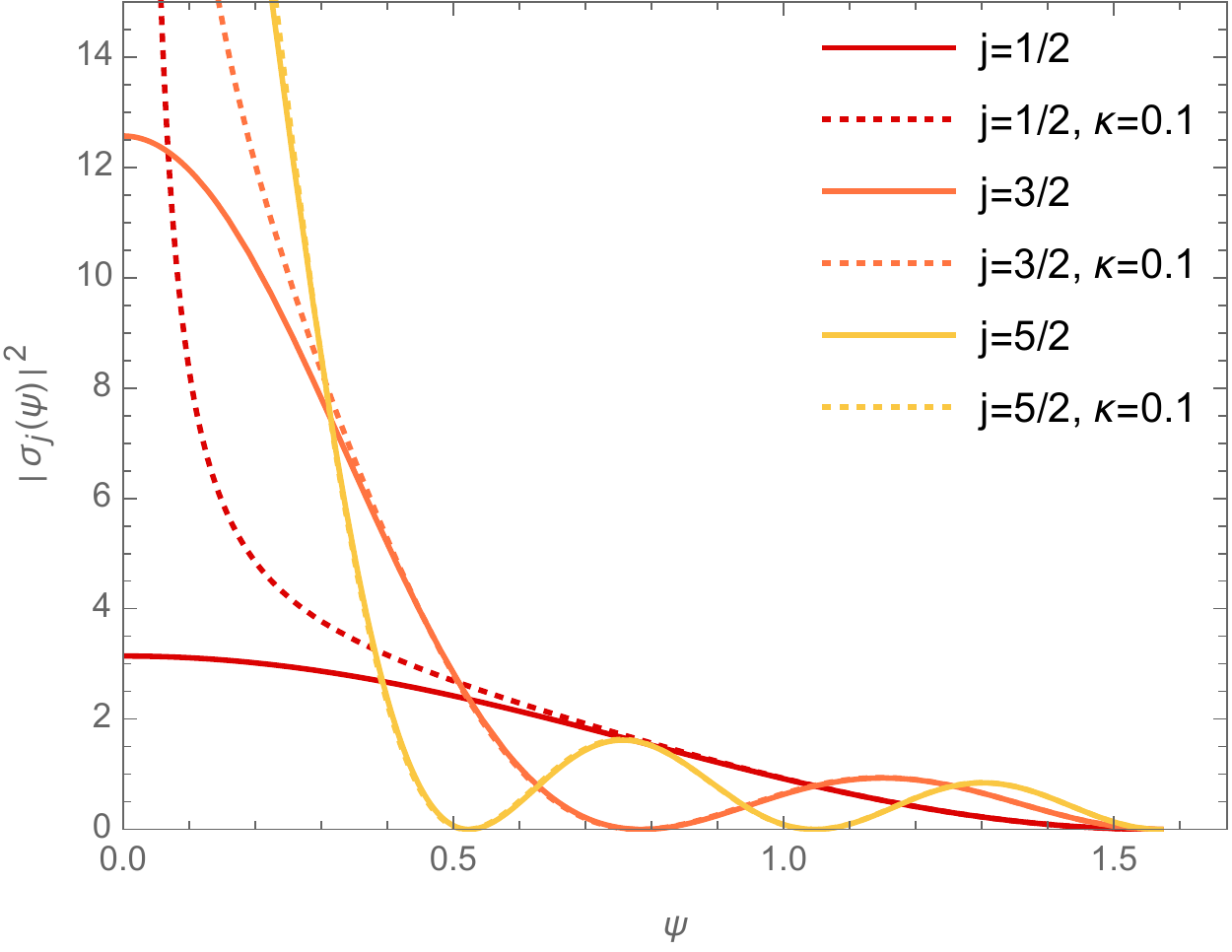}
	\caption{Probability density of the interacting mean field over $\psi$ for $V_{eff}[\sigma]=\frac{\mu}{2}\sigma^2+\frac{\kappa}{4}\sigma^4$.}
	\label{probdensityinteractingpsi}
\end{figure}

\begin{figure}[ht]
	\centering
   \includegraphics[width=0.4\textwidth]{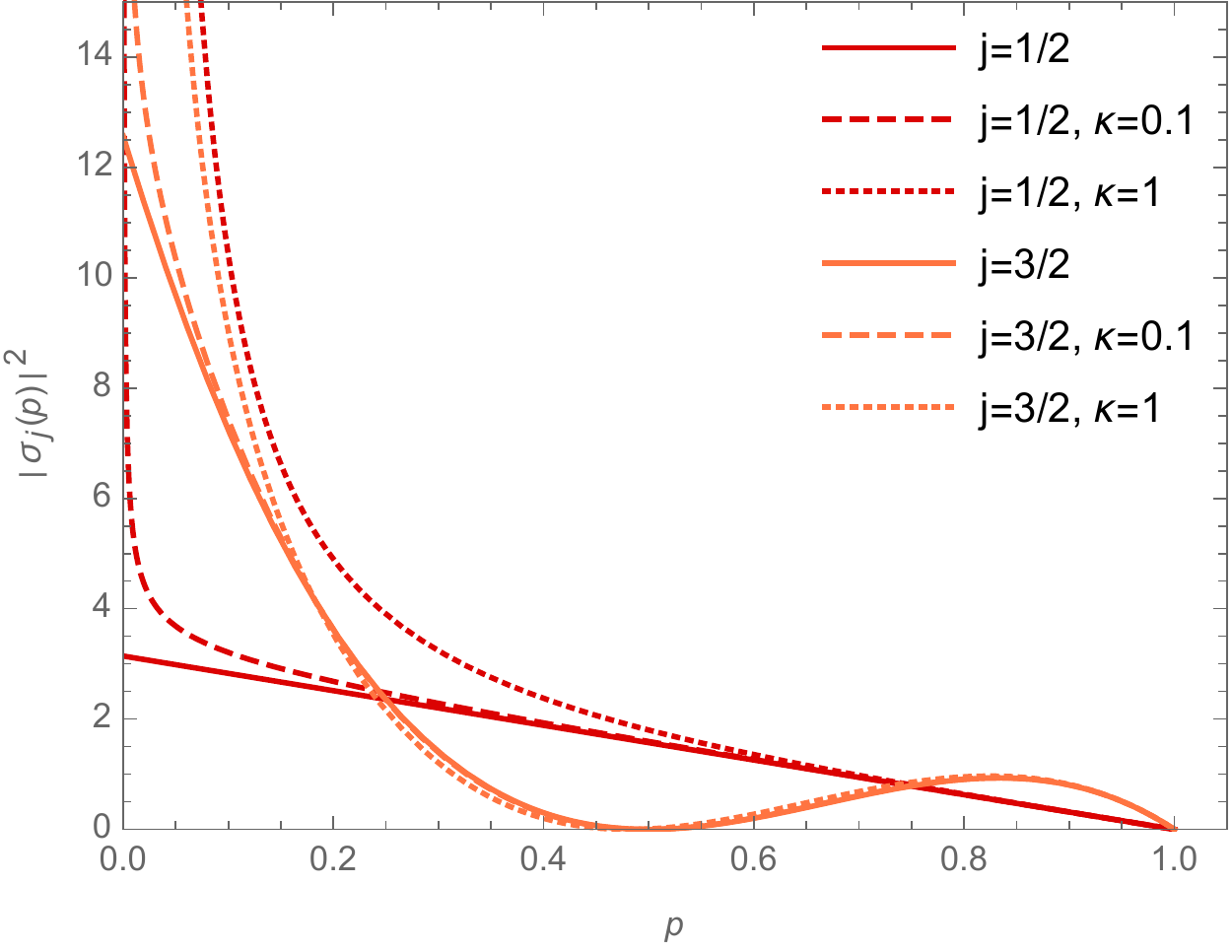}
	\caption{Probability density of the interacting mean field over $p$ for $V_{eff}[\sigma]=\frac{\mu}{2}\sigma^2+\frac{\kappa}{4}\sigma^4$.}
	\label{probdensityinteractingpvergleich}
\end{figure}

In the following, we discuss the behavior of solutions for the pseudotensorial potential (\ref{phi4potential}) with $\mu<0$ and $\kappa>0$ and where the qualitative results differ also for the pseudosimplicial potential (\ref{phi5potential}) with $\mu<0$ and $\kappa>0$ (or $\kappa<0$) so that the potentials would possess nontrivial minima. The effect of weak nonlinearities in the equation of motion onto the solutions is illustrated respectively in Fig. \ref{probdensityinteractingpsi} and Fig. \ref{probdensityinteractingpvergleich} in the $p$- and $\psi$-parametrizations and is contrasted to the behavior of the free solutions of subsection \ref{subsectionfree}. In general, the finiteness of the free solutions at the origin is lost due to the interactions. Crucially, the concentration of the probability densities around the origin can still be maintained giving rise to nearly flat solutions, as long as $|\kappa|$ does not become too big. For larger $j$, i.e. larger $|\mu|$, one sees that the departure from the free solutions is less pronounced because the $\mu$-term of the potential dominates longer over the $\kappa$-term. 

When $|\kappa|$ and $|\sigma'(1)|$ are small, solutions will remain normalizable with respect to the Fock space measure, i.e.
\be
N=\int (dg)^3 |\sigma(g_1,g_2,g_3)|^2<\infty.
\ee
However, when gearing up toward the strongly nonlinear regime, i.e. $\kappa\gtrsim\mathcal{O}(1)$, this feature is lost as $N$ grows and eventually one finds $N\to\infty$.\footnote{It should be noted that the precise values of $\kappa$ and/or $\sigma'(1)$ for which $N\to\infty$ depend on the numerical accuracy of the used solver. In this sense the observation of such behavior is a qualitative result.} The loss of normalizability of $\sigma$ with respect to the Fock space measure in the strongly nonlinear regime goes in hand with the breaking of the rescaling invariance expressed by Eq. (\ref{rescalinginvariance}) and signals the breakdown of the ansatz used here. Such behavior is not surprising, as it is well known within the context of local QFTs that the proper treatment of interactions necessitates the use of non-Fock representations for which $N$ is infinite (cf. appendix \ref{AppendixB} and \cite{FocknonFock}). We will get back to this point below.

With regard to the average of the field strength, one observes that $\kappa>0$ increases $\langle \hat{F}^i\rangle/N$ for some $j$ in comparison to the free case, whereas for negative $\kappa$ the expectation value decreases. This behavior is reminiscent of the effect of similar interactions onto the effective curvature of the space described by the condensate in Ref. \cite{GFCEmergentFriedmann3}, where it was shown that a bounded interaction potential generically leads to recollapsing condensate solutions.

\begin{figure}[ht]
	\centering
   \includegraphics[width=0.4\textwidth]{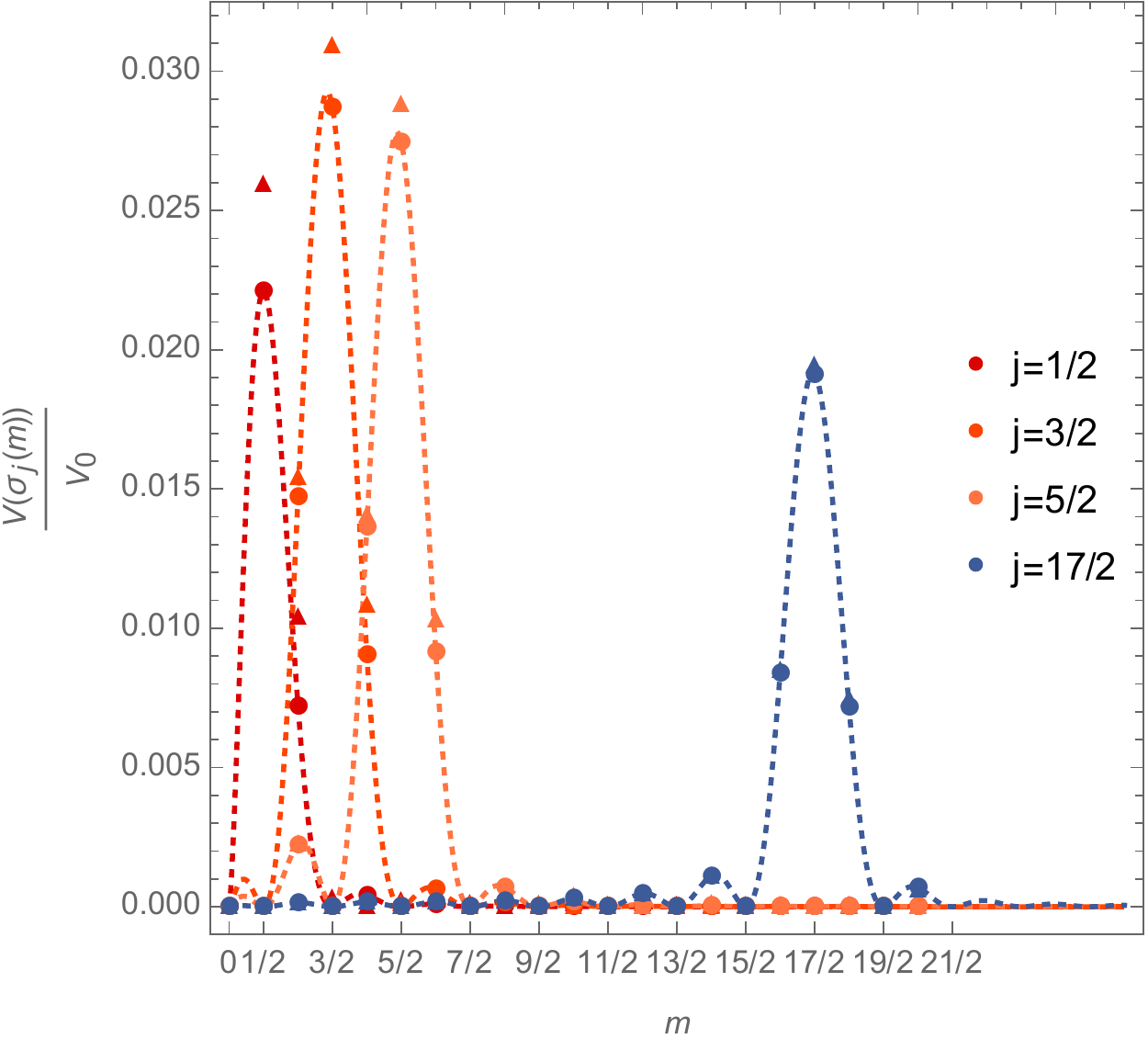}
	\caption{Normalized spectrum of the volume operator with respect to the interacting mean field $\sigma_j(\psi)$ for $\kappa=0.22$ (triangles) compared to the respective free solutions (dots).}
	\label{spectrumvolumeperturbed}
\end{figure}

\begin{figure}[ht]
	\centering
   \includegraphics[width=0.4\textwidth]{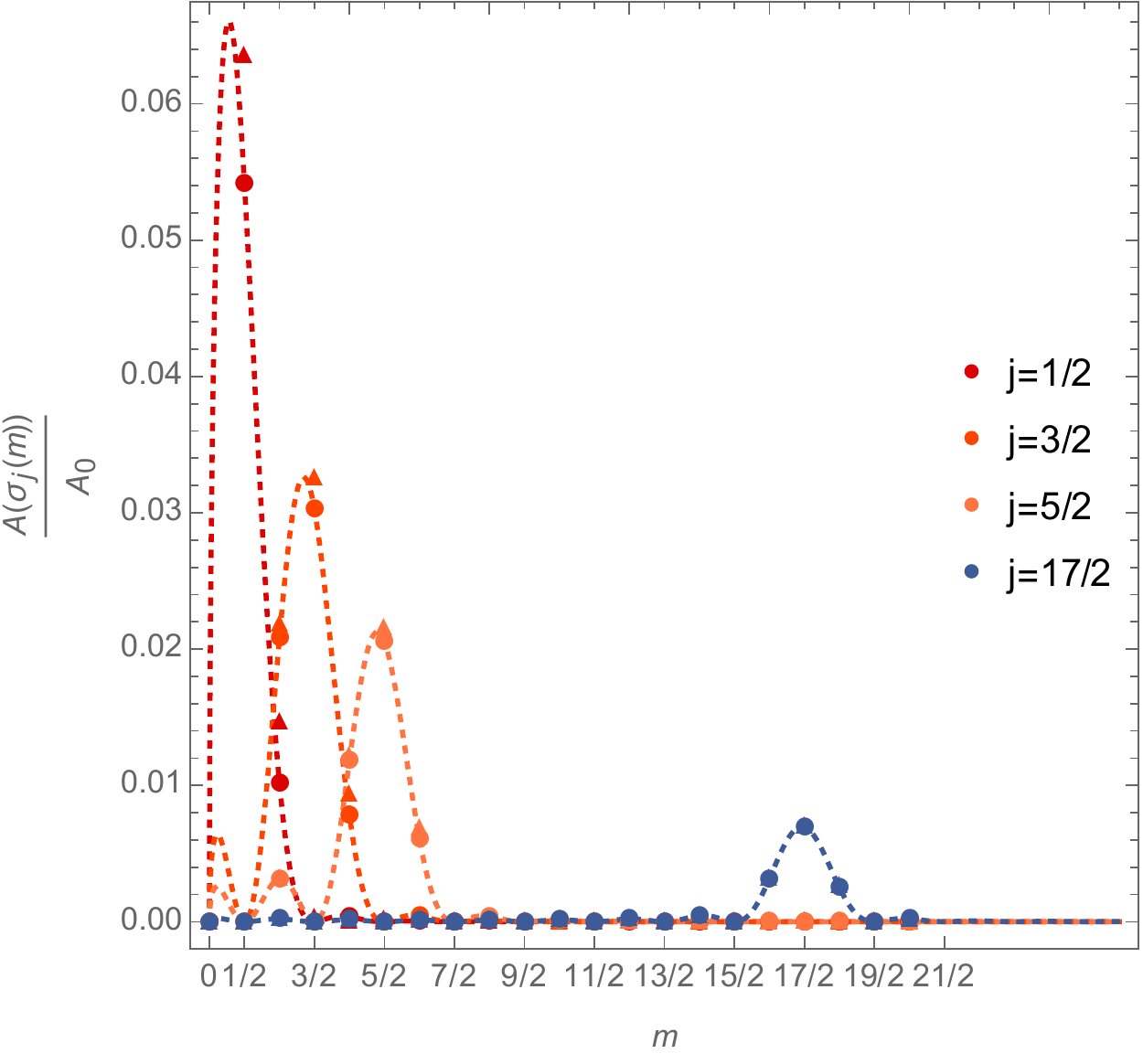}
	\caption{
Normalized spectrum of the area operator with respect to the interacting mean field $\sigma_j(\psi)$ for $\kappa=0.22$ (triangles) compared to the respective free solutions (dots).}
	\label{spectrumareaperturbed}
\end{figure}

By means of the numerically computed solutions, one can obtain their corresponding Fourier components and with these one yields in close analogy to the free case the modified spectra of the volume and area operators, illustrated in Fig. \ref{spectrumvolumeperturbed} and Fig. \ref{spectrumareaperturbed}. 
The plots clearly indicate that perturbations for $\kappa>0$ increase both the volume and the area, however, in the weakly nonlinear regime they remain finite. More specifically, one observes that the effect of the perturbations upon the spectra of the volume and area are more pronounced for small $j$, i.e. small $|\mu|$, since for these the nonlinearity dominates quickly over the $\mu$-term of the potential. Moreover, one notices that when pushing $\kappa$ to larger values as a consequence $V$ and $A$ quickly blow up in the same way as $N$ does, whereas $\langle \hat{F}^i\rangle/N$ remains finite. 

For the pseudosimplicial potential one obtains qualitatively analogous results with the differences to the free solutions being more emphasized since the nonlinearity is stronger.

\subsubsection{Solutions around the nontrivial minima}\label{subsectioninteractionslocalminima}

To chart the condensate phase and understand its properties, it is necessary to study numerically the solutions to the nonlinear differential equation (\ref{odesigma4}) around the nontrivial minima. To this aim, we choose for the coupling constants in Eq. (\ref{phi4potential}) in such a manner that the potential forms a Mexican hat, as in Fig. \ref{reellesfeldfi4}, and select the position of the minimum $\sigma_0$ as well as $\sigma'(1)$ as the boundary condition in order to find solutions numerically. 

Without any loss of generality, we will use the same values for $\mu$ as in the previous subsections. Apart from the requirement that they assume negative values they could be completely arbitrary since here we do not study eigensolutions to the Dirichlet Laplacian as in subsection \ref{subsectionfree}. 

\begin{figure}[ht]
\centering
  \hspace*{-0.85cm}\includegraphics[width=0.59\textwidth,height=3.5in]{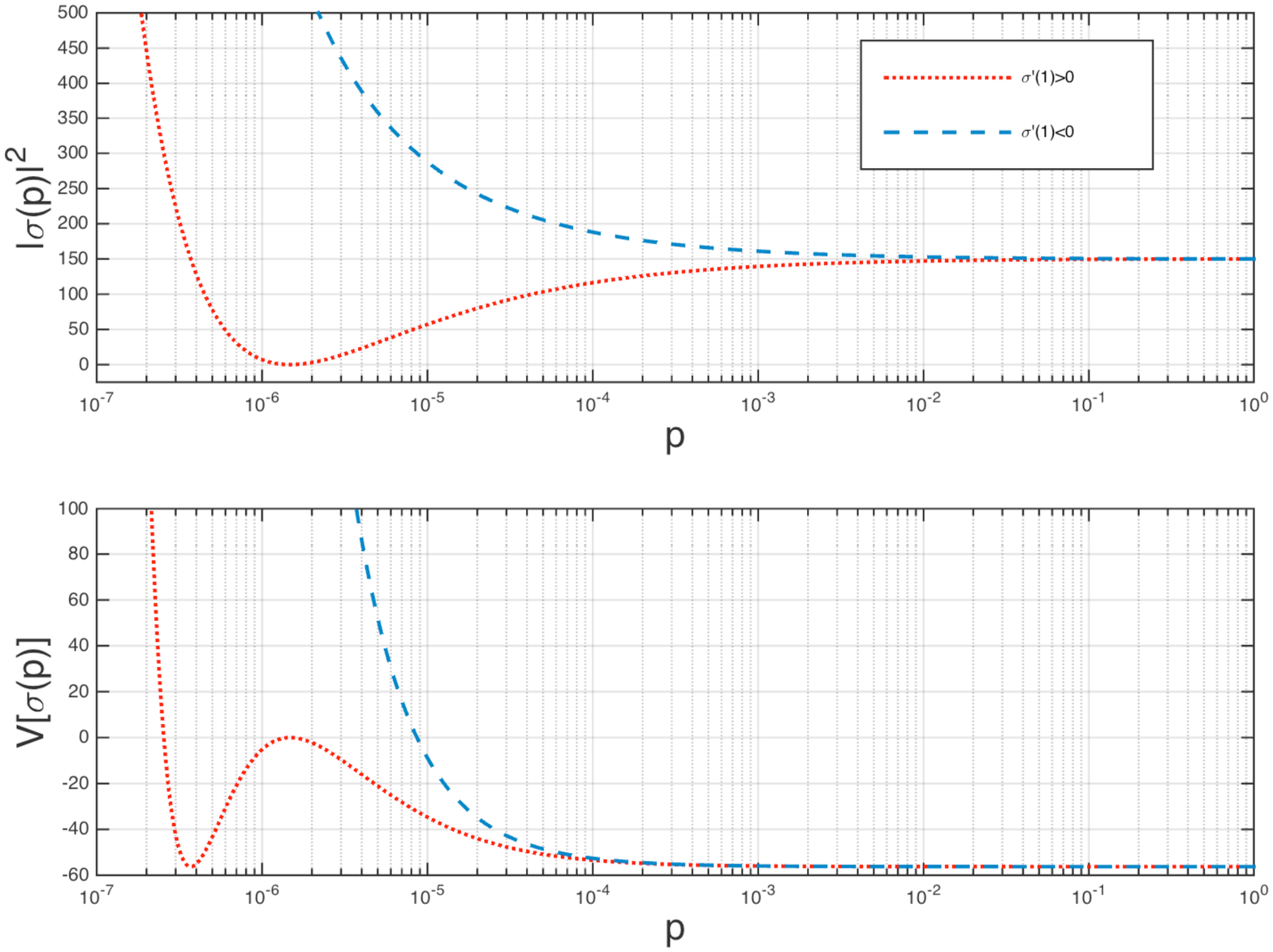}
  \caption{Semilog plot of the probability density and potential for solutions $\sigma(p)$ with $\mu=-1.5$, $\kappa=0.01$, $\sigma(1)=12.2474$ and $\sigma'(1)=\pm 100$ for the potential $V_{eff}[\sigma]=\frac{\mu}{2}\sigma^2+\frac{\kappa}{4}\sigma^4$. Solutions were computed by means of MATLAB's ODE45 solver which is based on an explicit Runge-Kutta $(4,5)$ formula. Output was generated for $10^5$ points on the interval $[0,1]$ while making use of highly stringent error tolerances.}
  \label{mexicanhatminimumsolutions}
\end{figure}

\begin{figure}[ht]
\centering
  \includegraphics[width=0.4\textwidth]{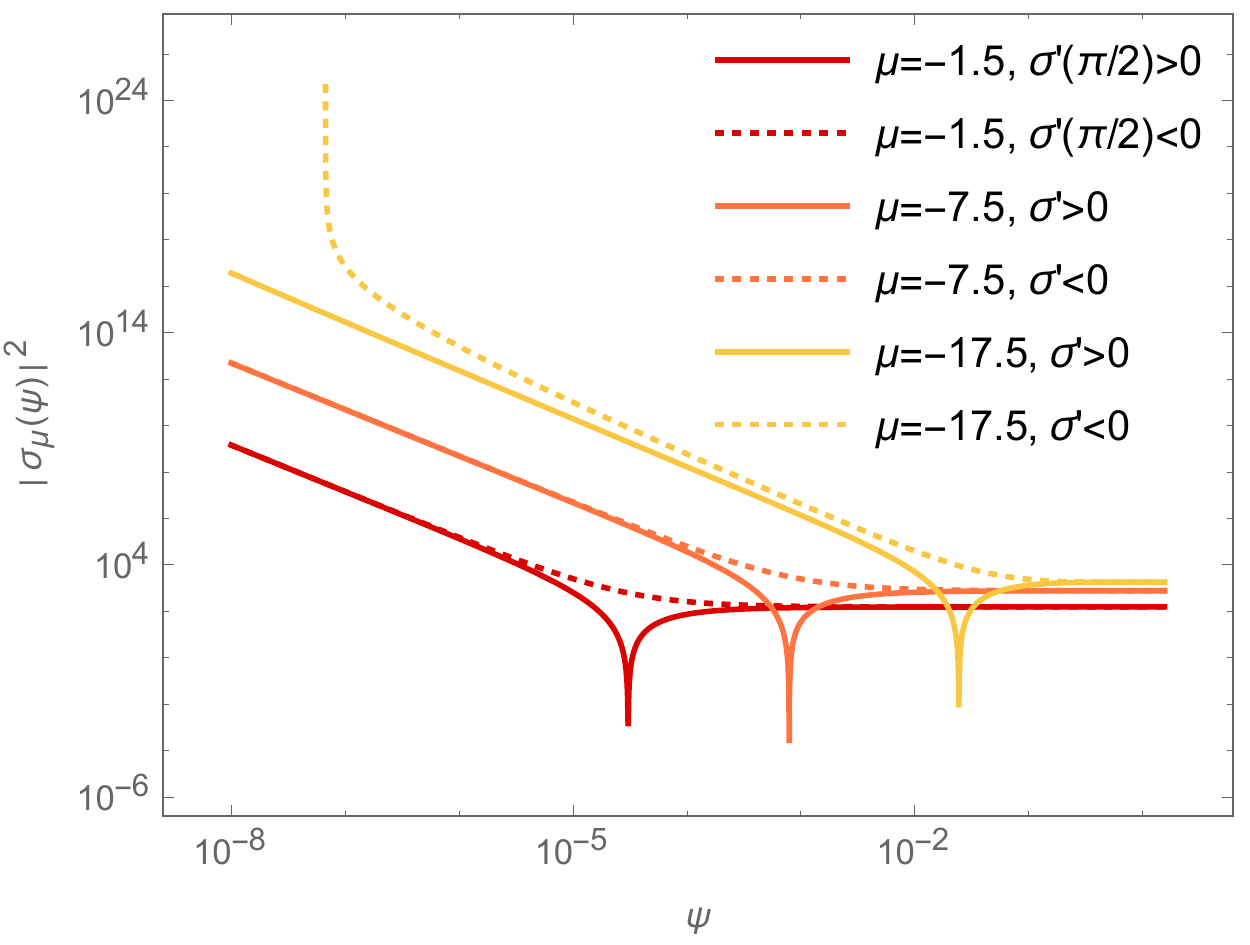}
  \caption{Double-log plot of the probability density for solutions $\sigma(\psi)$ for different $\mu$, with the same $\kappa=0.01$, and the same $|\sigma'(\pi/2)|$ at the respective minima $\sigma(\pi/2)$ for the potential $V_{eff}[\sigma]=\frac{\mu}{2}\sigma^2+\frac{\kappa}{4}\sigma^4$.}
  \label{mexicanhatminimumsolutionsloglog}
\end{figure}

Figure \ref{mexicanhatminimumsolutions} and Fig. \ref{mexicanhatminimumsolutionsloglog} show the resulting probability density and the potential over $p$ and $\psi$ computed for an exemplary choice for the values of the free parameters. Depending on the sign of $\sigma'(1)$ or $\sigma'(\pi/2)$ the solution either climbs over the local maximum at $\sigma=0$, then reaches the other minimum after which it ascends the left branch of the potential or directly climbs up the right branch shown in Fig. \ref{reellesfeldfi4}. For the choice of parameters leading to Fig. \ref{mexicanhatminimumsolutions} and \ref{mexicanhatminimumsolutionsloglog}, the solutions are normalizable. In general, for small $\sigma'(1)$ the solutions crawl slowly out of the minima and if $\sigma'(1)$ is almost zero, the solutions remain almost constant up to $p=0$, where the regular singularity of the differential equation finally kicks in. The contribution of the Laplacian term is less pronounced for smaller $\mu$ than for larger ones, as Fig. \ref{mexicanhatminimumsolutionsloglog} in the $\psi$-parametrization illustrates. Similar results are obtained when one keeps $\mu$ fixed while decreasing $|\sigma'(\pi/2)|$.


It is clear that as long as for the boundary condition $\sigma'\approx 0$ holds, this is equivalent to neglecting the Laplacian part of the kinetic term $\mathcal{K}$ in the equation of motion. The solutions, exemplified by Figs. \ref{mexicanhatminimumsolutions} and \ref{mexicanhatminimumsolutionsloglog}, show that the properties of the nontrivial ground state are then defined by the ultralocal action. Solving the equation of motion starting at the minima of the effective potentials gives rise to almost constant, i.e. homogeneous, functions on the domain.\footnote{Completely neglecting the Laplacian from the onset, is only justified when the interaction is dominant which corresponds to
the regime of large ground state condensate "density", i.e. $\kappa N \gg 1$. In the context of real BECs this is known as the Thomas-Fermi approximation \cite{BECs}.}

The geometric interpretation of such solutions which "sit" in the equilibrium position is slightly obstructed. This is due to the fact that the above-used near flatness condition cannot be straightforwardly applied to such solutions. Despite the fact that the probability density can be tuned to be concentrated around low curvature values, it is finite close to the equator at $p=1$, while in other cases it simply remains constant on the whole interval, as Fig. \ref{mexicanhatminimumsolutions} and Fig. \ref{mexicanhatminimumsolutionsloglog} show. This calls for a more differentiated formulation of this condition, perhaps by means of a well defined GFT-operator capturing the average curvature of the $3$-space described by means of the condensate state. 

\begin{figure}[ht]
	\centering
   \includegraphics[width=0.4\textwidth]{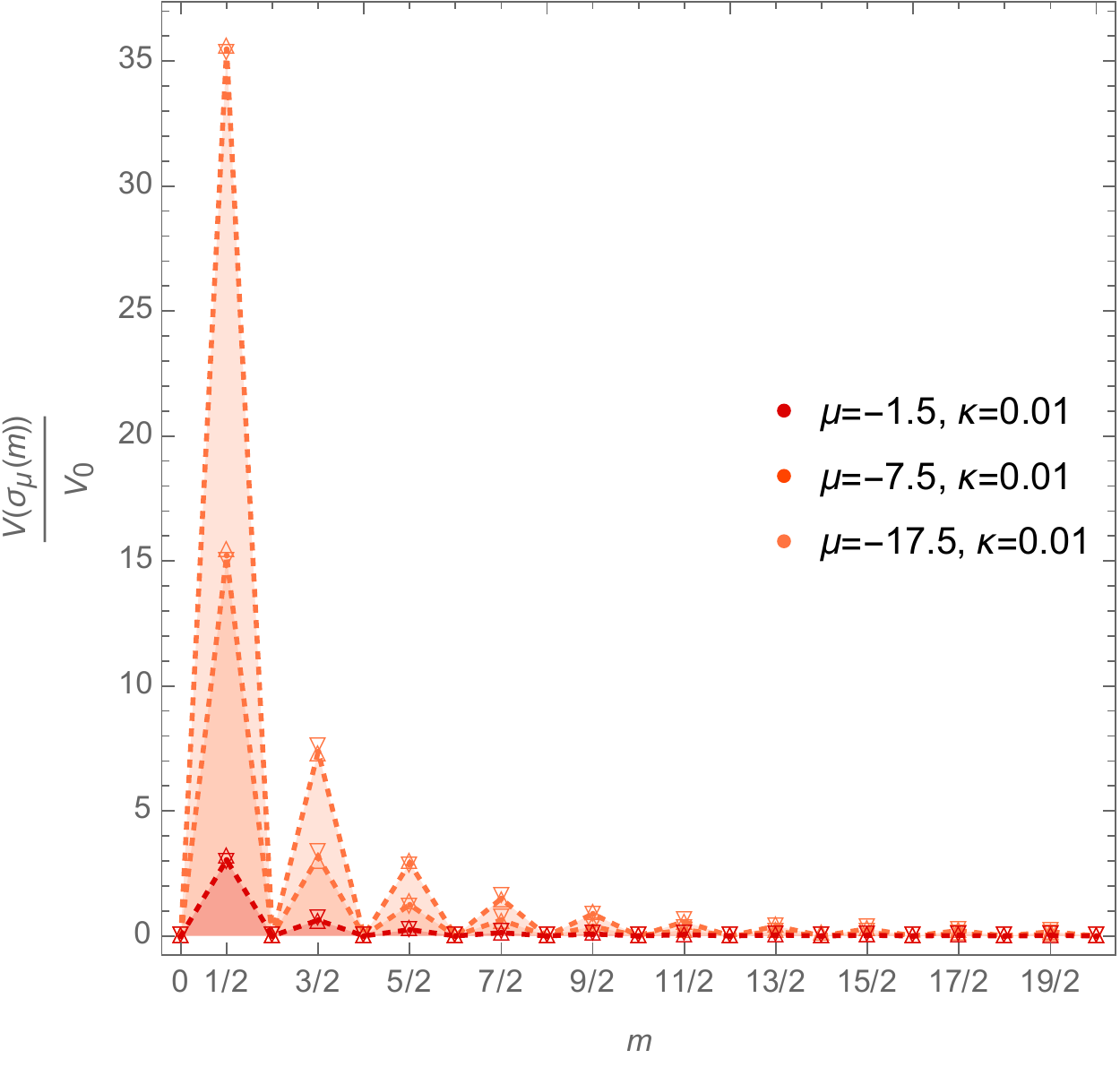}
	\caption{Normalized discrete spectrum of the volume operator (in arbitrary units) with respect to the interacting mean field $\sigma_{\mu}(\psi)$: Solutions $\sigma_{\mu}(\psi)$ were obtained with $\kappa=0.01$ but boundary conditions differ for each $\mu$ to solve around a nontrivial minimum of the respective potential $V_{eff}[\sigma]=\frac{\mu}{2}\sigma^2+\frac{\kappa}{4}\sigma^4$.}
	\label{spectrumvolumeperturbednontrivialminimum}
\end{figure}

\begin{figure}[ht]
	\centering
   \includegraphics[width=0.4\textwidth]{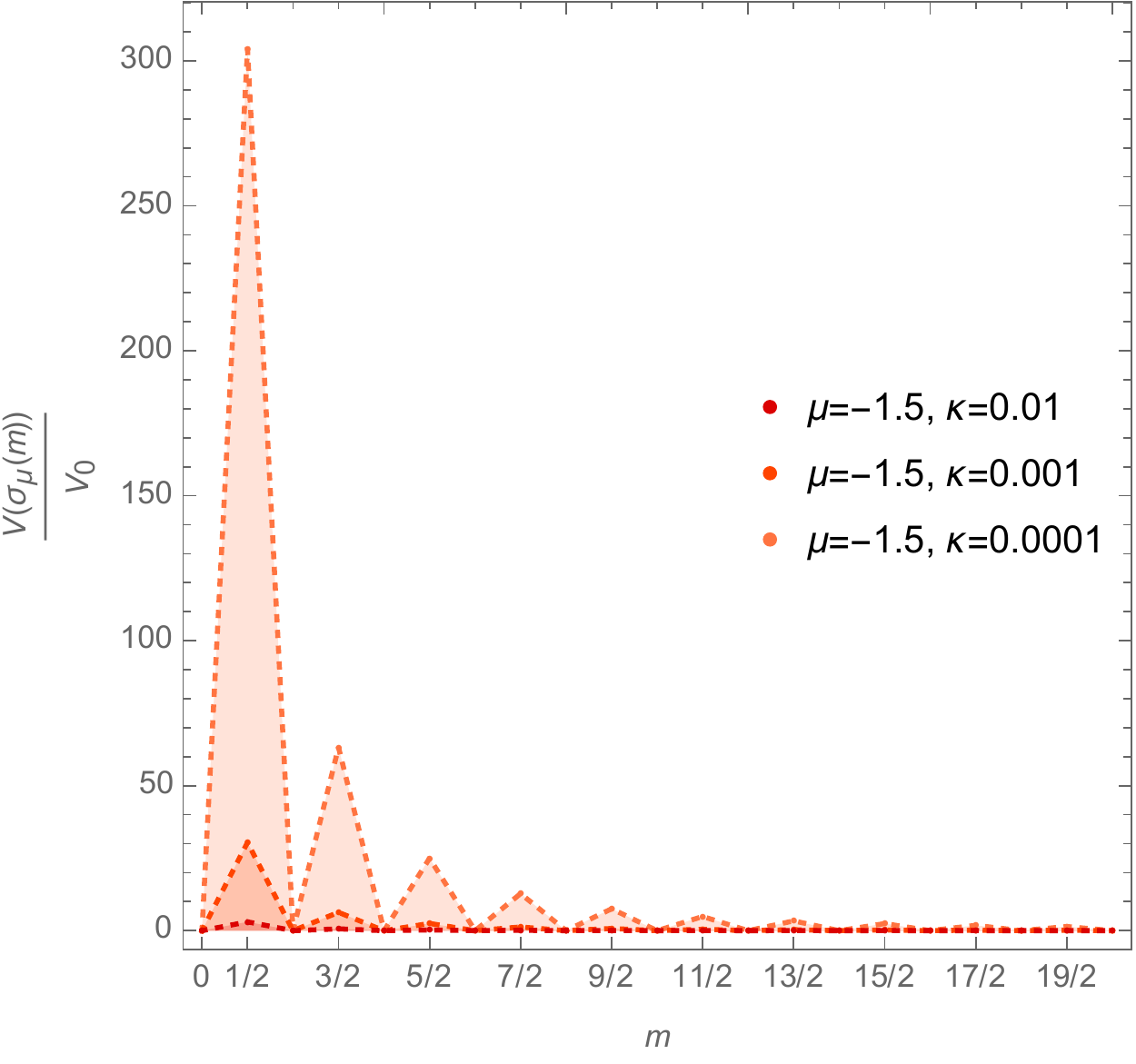}
	\caption{Normalized discrete spectrum of the volume operator (in arbitrary units) with respect to the interacting mean field $\sigma_{\mu}(\psi)$: Solutions $\sigma_{\mu}(\psi)$ were obtained for $\mu=-1.5$, different $\kappa$ and the same boundary conditions $\sigma'(\frac{\pi}{2})$ to solve around the respective nontrivial minima of the potential $V_{eff}[\sigma]=\frac{\mu}{2}\sigma^2+\frac{\kappa}{4}\sigma^4$.}
	\label{spectrumvolumeperturbednontrivialminimum2}
\end{figure}

\begin{figure}[ht]
	\centering
   \includegraphics[width=0.4\textwidth]{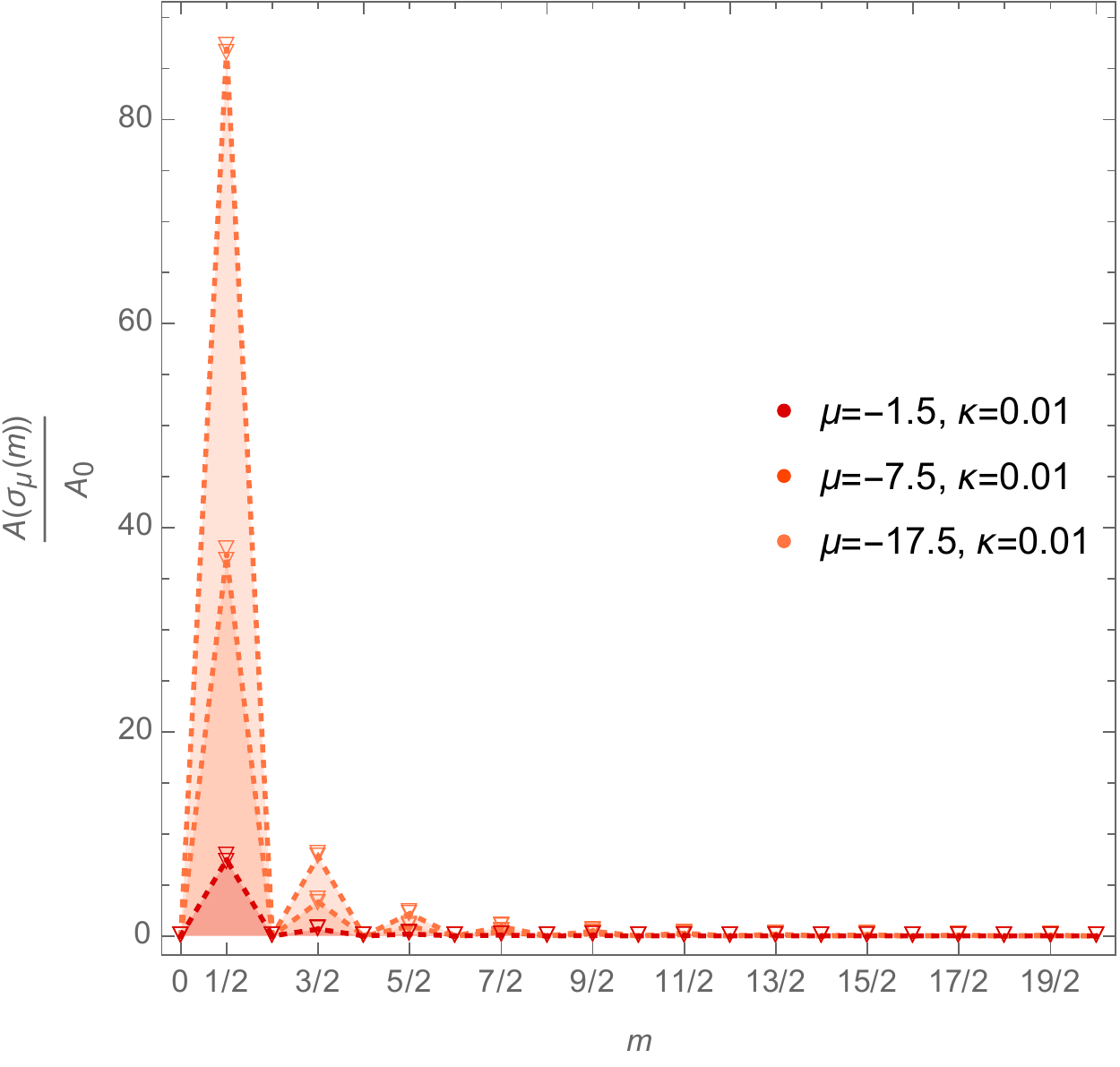}
	\caption{Normalized discrete spectrum of the area operator (in arbitrary units) with respect to the interacting mean field $\sigma_{\mu}(\psi)$: Solutions $\sigma_{\mu}(\psi)$ were obtained with $\kappa=0.01$ but boundary conditions differ for each $\mu$ to solve around a nontrivial minimum of the respective potential $V_{eff}[\sigma]=\frac{\mu}{2}\sigma^2+\frac{\kappa}{4}\sigma^4$.}
	\label{spectrumareaperturbednontrivialminimum}
\end{figure}

In spite of the current lack of such an operator to determine the curvature information stored in the mean field, it is possible to obtain from exemplary numerical solutions the spectrum of the volume and area operators as illustrated in Figs. \ref{spectrumvolumeperturbednontrivialminimum}, \ref{spectrumvolumeperturbednontrivialminimum2} and \ref{spectrumareaperturbednontrivialminimum}. Solutions which are computed around the nontrivial minima give rise to a different qualitative form of the spectrum of the volume and area as compared to the ones obtained in subsections \ref{subsectionfree} and \ref{subsectioninteractionsperturbation}; nevertheless we emphasize again the relevance of low spin modes. In general, for different $\mu$ the dominant contribution to the volume $V$ and area $A$ comes from the Fourier coefficients with $m=1/2$, whereas in the case discussed in the previous subsections the predominant contribution comes from the Fourier coefficients with $m=j$. This is due to the fact that $\sigma$ remains mostly constant and is thus best approximated by the simplest nontrivial mode for $m=1/2$. In particular, one can check that the contributions to $V$ and $A$ coming from the other modes, are exponentially suppressed when $\sigma'(1)\approx 0$.

Moreover, the volume and area remain finite in the weakly nonlinear case and when the boundary condition $\sigma'(1)$ is relatively small. The use of weak interactions is thus instructive in order to understand the qualitative behavior of the solutions in particular with regard to the expectation values of the geometric operators. Since the size of $\kappa$ has only a quantitative impact on the spectrum, as Fig. \ref{spectrumvolumeperturbednontrivialminimum2} suggests, an analogous form of the spectra can be expected also in the strongly nonlinear regime. Furthermore, for bigger values of $|\sigma'(1)|$ and/or strongly nonlinear interaction terms, the volume and area, as well as the expectation value of the number operator $\hat{N}$ blow up quickly. As noticed above, this signals the breakdown of the simple condensate state ansatz used here and suggests the need for non-Fock coherent states once the strongly correlated regime is explored (cf. appendix \ref{AppendixC}).\footnote{Figure \ref{spectrumvolumeperturbednontrivialminimum2} also seems to suggest that $V$ is ever increasing for $\kappa\to 0$. However, in such a limit, it is more appropriate to treat the system as in the free case, which we discussed in subsection \ref{subsectionfree}.}

Such solutions yield for all choices of $\mu<0$ and $\kappa>0$ for the averaged observables
\be
\frac{\langle\hat{\mathcal{O}}\rangle}{N}\approx \textit{const.},
\ee
since $\sigma$ is approximately constant. This naturally applies to the averaged field strength $\frac{\langle{\hat{F}^i\rangle}}{N}$ which is larger than in the corresponding free case. This indicates that the chosen effective GFT interactions have the effect of positively curving the effective geometry described by the condensate state. This is again reminiscent of similar findings in Ref.  \cite{GFCEmergentFriedmann3} where it was shown that relationally evolving and effectively interacting GFTCC models display recollapsing solutions when the interaction potential is bounded from below as here.

\begin{figure}[ht]
\centering
  \hspace*{-0.85cm}\includegraphics[width=0.59\textwidth,height=3.5in]{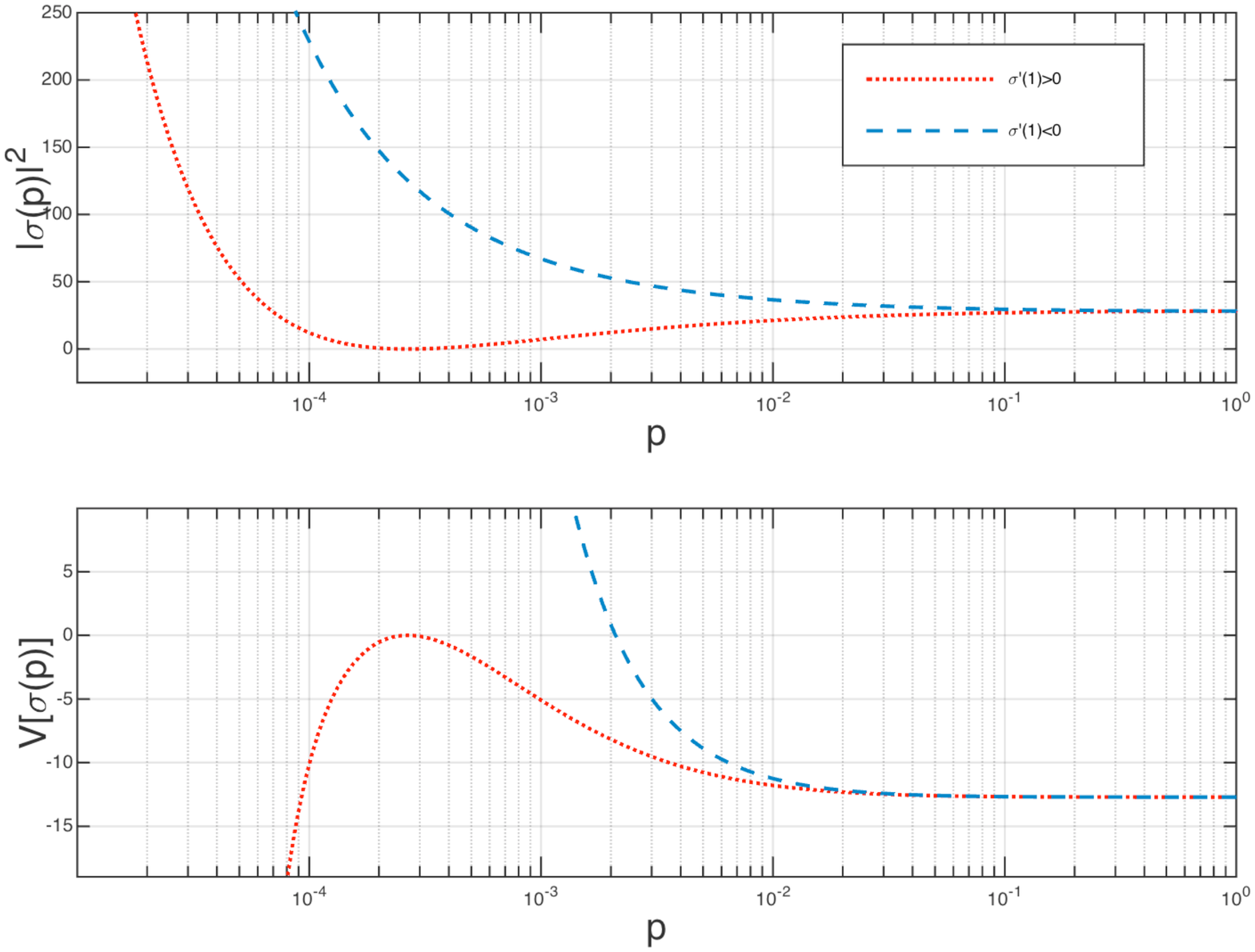}
  \caption{Semilog plot of the probability density and potential for solutions $\sigma(p)$ with $\mu=-1.5$, $\kappa=0.01$, $\sigma(1)=5.3132$ and $\sigma'(1)=\pm 100$ for the potential $V_{eff}[\sigma]=\frac{\mu}{2}\sigma^2+\frac{\kappa}{5}\sigma^5$. Solutions were computed by means of MATLAB's ODE113 procedure which is a variable order 'Adams-Bashforth-Moulton predictor-corrector' solver. Output was generated for $10^5$ points on the interval $[0,1]$ while making use of highly stringent error tolerances.}
  \label{pseudosimplicialminimumsolutions}
\end{figure}

Analogously, such a discussion can be repeated for the pseudosimplicial potential, where the solutions to the nonlinear equation of motion are illustrated in Fig. \ref{pseudosimplicialminimumsolutions}. The resulting behavior of the relevant operators is similar and will not be repeated here, though it should be kept in mind that only such interaction terms can be more closely related to models with a simplicial quantum gravity interpretation.

To summarize the main points of this subsection, we note that we have computed static condensate solutions around the nontrivial minima of the interaction potentials of which the essential features can be defined by means of the ultralocal action. We found that the condensate consists of many GFT quanta residing in the low spin mode $m=1/2$. This is indicated by the analysis of the discrete spectra of the geometric operators. Such low spins actually correspond to the IR regime of the theory. Hence, these results fit well into the picture suggested by the above- mentioned FRG analyses which find IR fixed points in all GFT models considered so far marking the formation of a condensate phase of which main features are supposed to be captured by means of the employed condensate state. In this sense, one can understand the condensate phase as to describe an effectively continuous homogeneous and isotropic $3$-space built from many small building blocks of the quantum geometry.

\section{Discussion and conclusion}\label{Discussion}

The purpose of this article was to investigate and interpret the impact of simplified interactions onto static GFT quantum gravity condensate systems describing effective $3$-geometries with a tentative cosmological interpretation. To this aim, we extensively examined the geometric properties of a free system in an isotropic restriction by studying the spectra of the volume and area operators imported from LQG and comparing the results to the perturbed case. In a last step, we studied the features of the GFT condensate when the system sits in the nontrivial minima of the effective interaction potentials. The main result of this study is then that the condensate consists of many discrete building blocks predominantly of the smallest nontrivial size encoded by the quantum number $m=1/2$ -- which supports the idea that an effectively continuous geometry can emerge from the collective behavior of a discrete pregeometric GFT substratum \cite{GFTGeometrogenesis}. In this sense, our results also strengthen the connection with LQC where the typically used quantum states are constructed from the assumption that the quanta of the geometry all reside in the same lowest nontrivial configuration \cite{LQC}. Together with the recently obtained results which show how free GFT condensate models dynamically reach a low spin phase \cite{GFClowspin}, this lends strong support to the idea that condensate states are appropriate for studying the cosmological sector of LQG.

The results of this article can also be seen as a support of the idea proposed in Ref. \cite{Carrozza}: The Laplacian in the kinetic operator $\mathcal{K}$, originally motivated by field theoretic arguments to guarantee the consistent implementation of a renormalization scheme, might only be a property of the UV completed GFT without a significant physical effect in the effectively continuous region which is expected to correspond to the small spin (IR) regime together with many building blocks of the quantum geometry. In this regime, the kinetic term is then suggested to become ultralocal, thus allowing for a straightforward interpretation of the GFT amplitudes in terms of spin foam amplitudes for quantum gravity. The numerical analysis done here indeed suggests that from the ultralocal action alone one can find that the condensate consists of many GFT quanta residing in the low spin configuration.

In the following we want to comment on the limitations of our discussion. We implicitly assumed that the condensate ansatz is trustworthy for any $\mu\leq 0$, where $\mu=0$ marks the critical value at which the phase transition from the unbroken into the condensate phase is supposed to take place \cite{TGFTFRG,GFTRGReview}. With respect to these findings, our analysis should be complemented by investigating whether indications for a phase transition into a condensate phase can be observed with the mean field techniques employed here, e.g., by means of the analyticity properties of the partition function, and whether their possible absence might be related to the expectation that true phase transitions are only realized for GFTs on noncompact manifolds, like Lorentzian quantum gravity models, as noticed in Ref. \cite{TGFTFRG}.

In this light, it is worth noting that in the context of weakly interacting, diluted and ultracold nonrelativistic BECs \cite{BECs} it is well understood that Bogoliubov's mean field and perturbation theory \cite{BogoliubovAnsatz} becomes invalid and breaks down in the vicinity of the critical point of the phase transition because quantum fluctuations become important. Of course, as is generally known today, mean field approaches work only accurately as effective descriptions of thermodynamic phases well away from critical points. A satisfactory description for such systems which systematically extends Bogoliubov theory and cures its infrared problems has been given in terms of FRG techniques \cite{WetterichFRGBEC}. The example of real BECs suggests that the analog of the Bogoliubov ansatz for quantum gravity condensates should be similarly extended by means of FRG methods at the critical point. This could be relevant for better understanding the nature of the phase transition which is possibly related to the "geometrogenesis" scenario.

It is also well understood that Bogoliubov theory for real BECs breaks down, when considering condensates with rather strongly interacting constituents. Likewise, FRG techniques can systematically implement nonperturbative extensions to Bogoliubov's approximation. These suggest that for Bose condensates with approximately pointlike interactions like in superfluid ${}^4\textrm{He}$, it is only possible to realize a strongly interacting regime for a very dense condensate \cite{WetterichFRGBEC}. This example could indicate a similar failure of the quantum gravity condensate ansatz when considering the strongly interacting regime. Indeed, when increasing the coupling constant $\kappa$ in this sector, the average particle number $N$ grows. If $\kappa$ is too large, we find that solutions are generally not normalizable with respect to the Fock space measure. The regime of large number of quanta $N$ and the eventual failure of $|\sigma\rangle$ to be normalizable in this sector certainly mark the breakdown of the Gross-Pitaevskii approximation to the dynamics (\ref{GPGFC}) for the simple condensate state constructed with Bogoliubov's ansatz (\ref{BogoliubovAnsatz}). In this regime, quantum fluctuations and correlations among the condensate quanta become relevant and only solutions to the full quantum dynamics together with FRG techniques would be capable of capturing adequately their impact. This entails that the approximation used here should only be trusted in a mesoscopic regime where $N$ is not too large, as already noticed in Ref. \cite{GFCEmergentFriedmann}. Nevertheless, the finding of solutions corresponding to non-Fock representations gives a forecast on what should be found when considering nonperturbative extensions of the techniques used here. 

In fact, the loss of normalizability is not too surprising because it is a generic feature of massless or interacting (local) QFTs according to Haag's theorem which require the use of non-Fock representations \cite{FocknonFock}. However, finding such solutions is first of all intriguing as a matter of consistency because non-Fock representations are also required in order to describe many particle systems in the thermodynamic limit. It is only in this limit that inequivalent irreducible representations of the CCRs become available which is a prerequisite for the occurrence of nonunique equilibrium states, in turn essential to consistently describing phase transitions \cite{FocknonFock}. It is also interesting for a second reason, since in the context of quantum optics it was understood that such non-Fock coherent states with an infinite number of (soft) photons can be described in terms of a classical radiation field \cite{nonFockCS}. Hence, the occurrence of non-Fock coherent state solutions in our context might also play a role in the classicalization of the system and could be important to consistently capture continuum macroscopic information of the GFT system. This would intuitively make sense, because one would expect to look for the physics of continuum spacetimes in the regime far from the perturbative Fock vacuum corresponding to the no-space state.

To fully extract the geometric information encoded by such solutions, it would then also be necessary to go beyond the use of the simplified local interactions and explore the effect of the proper combinatorially nonlocal interactions encountered in the GFT literature (e.g. on the expectation values of the geometric operators) in order to compare it to the results obtained here. Additionally, it is worth mentioning that only for proper simplicial interaction terms the quantum geometric interpretation is rather straightforward while for the others a full geometric interpretation is currently lacking.

In a next step, the time evolution of the condensate with respect to a relational clock could be studied. This would be in the spirit of \cite{GFCEmergentFriedmann}, where for an isotropic and free condensate configuration a Friedmann-like evolution was found. It would allow for the comparison between the two settings and the extraction of further phenomenological consequences from our model.

Another, perhaps more consistent way to properly relate the quantum geometric information stored in the condensate to a classical counterpart, could be to reconstruct the metric from the mean field as reviewed in appendix \ref{AppendixA} and investigate its isometries. Should the condensate approximate a continuous homogeneous geometry, one would expect that diffeomorphism invariance be restored as highlighted in a related context in Ref. \cite{Dittrich}. This could be helpful when comparing the isotropic restriction employed in Ref.  \cite{GFCEmergentFriedmann} to the one used here and also when anisotropic condensates are studied. 

Finally, we remarked in our analysis that the notion of near flatness used in the previous subsections should be reconsidered for the condensate solutions around the nontrivial minima since it cannot be straightforwardly applied then. Statements regarding the flatness property of such solutions can only be satisfactorily made if the spectrum of the currently unavailable GFT curvature operator is studied. Such possible extensions will be explored elsewhere.
{\bf Acknowledgements.} AP is grateful to L. Sindoni, S. Gielen, D. Oriti, J. Th\"urigen, S. Carrozza, M. De Cesare and M. Schwoerer for helpful remarks.

\begin{appendix}

\section{Noncommutative Fourier transform and reconstruction of the metric}\label{AppendixA}

The following discussion reviews how GFT states can be understood to encode quantum geometric information dressing $3d$ simplicial complexes and thus express how spatial slices can be triangulated. For a detailed discussion we refer to Refs. \cite{GFC,GFCReview}.

As is well known, in the Hamiltonian formulation of Ashtekar-Barbero gravity, where $G=\textrm{SU}(2)$, the canonically conjugate variable to the gravitational connection is given by the densitized inverse triad. From these momentum space variables the spatial metric can be derived, making the geometric interpretation perhaps more transparent. Motivated by this, we want to reformulate the GFT formalism in terms of these variables by means of a noncommutative Fourier transform (ncFT) which allows us to shift in between configuration and momentum space \cite{NCFT}.

To this aim, let $G^d$ with $d=4$ be the configuration space of the GFT field, then the phase space is given by the cotangent bundle $T^{*}G^4\cong G^4\times\mathfrak{g}^4$. The ncFT of a square integrable GFT field is then given by
\be\label{ncFT}
\hat{\tilde{\varphi}}(B_I)=\int (dg)^4\prod_{I=1}^4 e_{g_{I}}(B_I)\hat{\varphi}(g_I),
\ee
wherein the fluxes $B_I$ with $I=1,...,4$ parametrize the noncommutative momentum space $\mathfrak{g}^4$ and $e_{g_{I}}(B_I)$ is a choice of plane waves on $G^4$. Their product is noncommutative, i.e., $e_{g}(B)\star e_{g'}(B)=e_{gg'}(B)$, signified by the star product. By means of the noncommutative Dirac delta distribution in the momentum space representation
\be
\delta_{\star}(B)=\int dg~ e_{g}(B),
\ee
it can be shown that the invariance of the GFT fields under the right diagonal action of $G$ yields a closure condition for the fluxes, i.e., $\sum_I B_I=0$. It guarantees the closure of $I$ faces dual to the links $e_I$ to form a tetrahedron. It also allows us to eliminate one of the $B_I$s when reexpressing the fluxes in terms of discrete triads. This is done by $B_i^{ab}=\int_{\triangle_i}e^a \wedge e^b$ with the cotriad field $e^a\in\mathbb{R}^3$ encoding the simplicial geometry and $i=1,2,3$ associated to the faces $\triangle_i$ of the tetrahedron. From this, the metric at a given fixed point in the tetrahedron can be reconstructed leading to
\be\label{reconstructedmetric}
g_{ij}=e^a_i e^b_j\delta_{ab}=\frac{1}{4 \tr(B_1 B_2 B_3)}\epsilon_i^{kl}\epsilon_{j}^{mn}\tilde{B}_{km}\tilde{B}_{ln},
\ee
with $\tilde{B}_{ij}\equiv \tr(B_iB_j)$ \cite{GFC}. In this way, $\hat{\tilde{{\varphi}}}^{\dagger}(B_i)|\emptyset\rangle=|B_i\rangle$ determines the metric of a quantum tetrahedron. Notice that for the condensate mean field $\sigma(g_I)$ the ncFT can be straightforwardly computed by means of (\ref{ncFT}) and the reconstruction of the metric (\ref{reconstructedmetric}) holds then for all constituents of the condensate.

\section{Fock and non-Fock representations}\label{AppendixB}
Following Ref. \cite{GFTLQG}, for the Fock representation of GFT one defines a set of fundamental operators $\hat{c}_i$ and $\hat{c}^{\dagger}_i$, with the algebraic relations
\be\nonumber
[\hat{c}_i,\hat{c}^{\dagger}_{i'}]=\delta_{ii'}~~\textrm{and}~~[\hat{c}^{(\dagger)}_i,\hat{c}^{(\dagger)}_{i'}]=0
\ee
satisfying 
\be\nonumber
\hat{c}_i |N_i\rangle=\sqrt{N_i} |N_i-1\rangle~~\textrm{and}~~\hat{c}^{\dagger}_i |N_i\rangle=\sqrt{N_i+1} |N_i+1\rangle.
\ee
The operators $\hat{c}_i$ and $\hat{c}_i^{\dagger}$ annihilate and create single spin network vertices acting on the Fock vacuum state given by
\be\nonumber
\hat{c}_i|\emptyset\rangle=0,~~\forall i,
\ee
with the single-vertex label $i$ characterizing the quantum geometric properties of the state. The occupation number operators are then expressed by
\be\nonumber
\hat{N}_i|N_i\rangle=\hat{c}_i^{\dagger}\hat{c}_i|N_i\rangle=N_i|N_i\rangle
\ee
with the total number operator $\hat{N}=\sum_i \hat{N}_i$. 

Within the context of local QFT \cite{FocknonFock} it is well known, that in the finite dimensional and noninteracting infinite dimensional cases all irreducible Fock representations are unitarily equivalent and hence there is just one phase associated to the quantum system. However, this is different for interacting fields, models with nonvanishing ground state expectation value and many body systems in the thermodynamic limit, where the Fock representation is not allowed and $N$ is not a good quantum number for the characterization of the system since $\hat{N}$ is unbounded from above. In these situations the systems are described by means of non-Fock representations corresponding to inequivalent representations of the commutation relations and thus allow for the occurrence of different phases associated to the considered quantum system. Though these statements currently lack an axiomatic underpinning from within the GFT context, we believe that their basic intuition also holds there.

\section{Non-Fock coherent states}\label{AppendixC}

In the following we clarify the notion of a non-Fock coherent state following largely the established literature on optical coherence in Refs. \cite{FocknonFock,nonFockCS} and try to link it to the GFT formalism. To this aim, we introduce some axiomatic terminology.

From an algebraic point of view, it is known that a quantum system is defined by its algebra of observables $\mathcal{A}$ being a unital $C^{*}$-algebra. A state is a linear functional $\omega:\mathcal{A}\to \mathbb{C}$ which is positive (i.e. $\omega(a^{\dagger}a)\geq 0~\forall a\in\mathcal{A}$) and normalized (i.e. $\omega(\mathbb{1})=1$) with $\omega(A)=\langle A\rangle$. Without proof let us assume that for each such $\omega$ there is a GNS triple (determined up to unitary transformations), $(\mathcal{F}_{\omega},\pi_{\omega},\psi_{\omega})$, where $\mathcal{F}_{\omega}$ is the bosonic Fock space, $\pi_{\omega}$ is a unit-preserving representation of $\mathcal{A}$ in terms of linear operators over $\mathcal{F}_{\omega}$ and $\psi_{\omega}\in\mathcal{F}_{\omega}$ is cyclic, that means $\pi_{\omega}(\mathcal{A})\psi_{\omega}$ is dense in $\mathcal{F}_{\omega}$. Using the scalar product in $\mathcal{F}_{\omega}$, $\langle \psi_{\omega}|\pi_{\omega}(a)\psi_{\omega}\rangle=\omega(a)$ holds for all $a\in \mathcal{A}$. Using this language, in relation to appendix \ref{AppendixB} one can write for example $\langle N_i\rangle=\omega(\hat{c}^{\dagger}_i \hat{c}_j)=N_i$.

In the following, let the domain $\mathcal{C}=\mathrm{SU}(2)\backslash \mathrm{SU}(2)^4/\mathrm{SU}(2)$ and $dh$ denotes the measure on $\mathcal{C}$ with $g_I\in\mathcal{C}$. Using the distributional character of the field operators, we smear the creation and annihilation operators with the real functions $f_i\in C_0^{\infty}(\mathcal{C})$ which form an orthonormal set $\{f_i\}$, giving e.g.  $\hat{c}(f_i)=\hat{\psi}(f_i)=\int_{\mathcal{C}} dh~\hat{\psi}(g_I)f_i(g_I)$.

Using the above, a state $\omega$ is called (fully) coherent if it possesses a factorization property of the correlation functions in the sense that with a linear form, the coherence function, $L:C_0^{\infty}(\mathcal{C})\to\mathbb{C}$ one has
\begin{multline}\nonumber
\omega(\hat{c}^{\dagger}(f_1)\cdots\hat{c}^{\dagger}(f_k)\hat{c}(g_1)\cdots\hat{c}(g_k))=~\\L(f_1)\cdots L(f_k)\bar{L}(g_1)\cdots\bar{L}(g_l)
\end{multline}
for all $k,~l\in\mathbb{N}_0$ with $k=l$ and for all $\{f_k\}$ and $\{g_l\}\in C_0^{\infty}(\mathcal{C})$. In particular, $\omega(\hat{c}^{\dagger}(f_i)\hat{c}(f_i))=|L(f_i)|^2\stackrel{!}{=}N_i$ holds. One calls the coherence function $L$ bounded, if there exists a constant $c_L\geq 0$ with $|L(f)|\leq c_L ||f||$. Otherwise $L$ is unbounded. In our context, $f$ is strictly related to the mean field $\sigma$.

With this one can make the following statements. A coherent state $\omega$ is normal to the Fock representation, if and only if $L$ is bounded, that means the state is given by a unique density operator in Fock space. For unbounded $L$ the state $\omega$ is not representable by a density operator in Fock space, i.e., $\omega$ is disjoint from the Fock sector. This implies that the set of all occupation numbers is unbounded.

Suppose now, that $\omega$ is a coherent state in the above sense. For bounded $L$ one calls $\omega$ a Fock coherent or a microscopic coherent state. In contradistinction to that one calls $\omega$ a non-Fock coherent or a macroscopic coherent state if the coherence function $L$ is unbounded. One can show that $L$ exhibits then specific classical features, such as a collective phase and amplitude which means that it acquires the status of a classical field due to the ordering effect of the present phase correlations. Furthermore, one can show that the unboundedness of $L$ leads to a finite particle density in the infinite volume in contrast to a vanishing particle density for bounded $L$ in the same limit \cite{nonFockCS}. Despite the fact, that these statements currently lack a rigorous underpinning from within the GFT context, again, we believe that their intuition could be directly transferred.

\end{appendix}

\end{document}